\documentclass[twocolumn,preprintnumbers,amsmath,amssymb]{revtex4}
\usepackage{tabularx,graphicx}

\usepackage{color}
\usepackage{hyperref}
\hypersetup{
    colorlinks=true,
    linkcolor=blue,
    filecolor=blue,      
    urlcolor=blue,
}

\usepackage{color}

\usepackage{ulem}   

\begin{document}

\newcommand{\beq}{\begin{equation}}
\newcommand{\eeq}{\end{equation}}
\newcommand{\beqn}{\begin{eqnarray}}
\newcommand{\eeqn}{\end{eqnarray}}
\newcommand{\bmath}{\begin{subequations}}
\newcommand{\emath}{\end{subequations}}
\newcommand{\bra}[1]{\langle #1|}
\newcommand{\ket}[1]{|#1\rangle}

\title{The Meissner effect in superconductors: emergence versus reductionism}

\author{J. E. Hirsch}
\address{Department of Physics, University of California, San Diego,
La Jolla, CA 92093-0319}
 
 \begin{abstract} 
The Meissner effect, the expulsion of magnetic field from the interior of a metal entering the superconducting state,
is arguably the most fundamental property of superconductors, discovered in 1933. 
The conventional theory of superconductivity developed in 1957 is generally believed to fully explain the Meissner effect.
We will review the arguments that support this consensus, rooted in the concept of emergence. However,  recent work has shown that there are  questions
related to momentum conservation in the process of magnetic field expulsion that have not  been addressed within the conventional
theory. Within a  reductionist approach, it has been proposed that those questions can only   be resolved by introducing physics
that is not part of the conventional theory, namely that there is  {\it radial motion of electric charge}
in the transition process. This is  consistent with the behavior of classical plasmas, where motion of magnetic field lines
is always associated with motion of charges.   We review how this approach explains puzzles associated with
momentum transfer between electrons and ions in the Meissner effect. 
 Whether or not radial charge motion is associated with the
 Meissner effect has fundamental implications regarding  superconductivity mechanisms in materials  and regarding strategies to
 search for new materials with higher superconducting transition temperatures.
 Therefore, adjudication of this question is urgent and important. \end{abstract}
 \maketitle 
 
\tableofcontents

\section{Introduction}
The two most characteristic properties of superconductors are zero resistance and the Meissner effect, the expulsion of magnetic fields from
the interior of a material becoming superconducting. The Meissner effect is the
property that differentiates superconductors from ``perfect conductors'', which are  materials that conduct electricity
with no resistance but do not expel magnetic fields. Perfect conductors do not exist, but can be regarded as a limiting case
of ``very good conductors'', that do exist. What makes superconductors $qualitatively$ different from very good or perfect conductors is precisely the Meissner
effect. When the effect was  experimentally discovered in 1933 \cite{meissner} it triggered a very rapid advance in the field, because it was completely unexpected
and because  it is such a fundamental property. We can reasonably expect that such a fundamental property of superconductors
should reflect fundamental physics underlying the superconductivity phenonenon. What that fundamental physics is is currently under contention and is the subject of this review.


 It is generally believed in the scientific community that   the Meissner effect  is explained by the conventional theory of superconductivity, BCS \cite{bcspaper,tinkham,degennes}.
And even for unconventional theories proposed to describe what are believed to be ``unconventional superconductors'' \cite{unconventional} not described
 by BCS, it is believed that the Meissner effect is explained by the same principles that explain it in the conventional theory. 
 This conventional understanding of the Meissner effect characterizes it as an emergent property of the systems \cite{emergence,emergence2}, and does not 
 address in detail processes and mechanisms involved in the Meissner effect. In essence, it postulates that the physical systems
 will find ways to reach their lowest energy states. Instead, in recent years an alternative reductionist approach has emerged \cite{transitionprocess},
 which asks and proposes detailed answers to how phenomena associated with the Meissner effect occur, and requires specific properties
 of the physical systems, in the absence of which  the Meissner effect would not take place.

 
The basic principles that govern the Meissner effect that I will discuss here within both alternative points of view should be understandable to
a broad audience of physicists with no specialized background in the theory of superconductivity. Which one is correct, if any, is unknown,
and has fundamental implications for the future advancement in the understanding of superconductivity and its practical applications.

%


\section{What the Meissner effect is and is not}
It is often said that the Meissner effect is the $exclusion$ of magnetic fields from the interior of superconducting materials \cite{mahan}. 
It is also  often said, in describing  experiments where  a  material is
first cooled into the superconducting state and then a magnetic field is applied, that the  superconductor $expelled$ the magnetic field \cite{eremets}. Both
statements  are  
wrong, and it is worth emphasizing it here at the outset because these misconceptions are pervasive in the literature  and muddle the essential issue.
Certainly the conventional theory of superconductivity can explain the Meissner effect as described in this paragraph. But that is $not$ the Meissner
effect.

The Meissner effect is $not$ the state of a superconductor, even if superconductors with magnetic field excluded are often said to be in the ``Meissner state''.
The Meissner effect (Fig. 1)  is the $process$ of  $expulsion$ of a pre-existent magnetic field from the interior of a metal that is initially in the normal state
and transits into the superconducting state,
that occurs either when the temperature is lowered or the external magnetic field is lowered. 
In this paper we will consider only type I superconductors, where the expulsion of magnetic field occurs suddenly at
given values of temperature and magnetic field. Instead, in type II superconductors the expulsion of magnetic fields occurs over
a finite range of temperature or magnetic field. In this paper we want to understand the Meissner $effect$: 
what is the physics that gives rise to the $process$ whereby the material goes from its initial normal state with magnetic field in the interior
to its final superconducting state with magnetic field excluded. To understand the process it is not sufficient to 
describe the equilibrium properties of the superconductor in no matter how much detail, as is often done.

         \begin{figure} [t]
 \resizebox{8.5cm}{!}{\includegraphics[width=6cm]{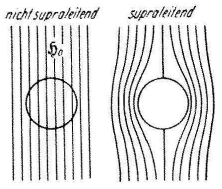}} 
 \caption { Graphic depiction of the Meissner effect from Meissner and Heidenreich's 1936 paper \cite{meissner1936}.}
 \label{figure1}
 \end{figure}

\section{The Meissner-Ochsenfeld discovery, and why it took so long}
The discovery of the Meissner effect was entirely accidental and entirely unexpected.
Meissner and Ochsenfeld in early 1933 designed an experiment to test a calculation performed by Laue and M\"oglich \cite{laue} on the expected change in 
the current distribution in a wire transitioning  between normal and superconducting states. To do so, they circulated current in two parallel wires in opposite direction
and measured the magnetic field in the region between the wires. In one of several measurements they found the surprising result that the measured magnetic field was
different when the directions of the currents were switched. This was interpreted by Meissner to imply that the difference originated in the earth's magnetic
field, that had not been compensated for in that particular measurement, and hence indicated  that a change in magnetic field in the neighborhood of the wires would also occur
in the absence of applied current \cite{meissner1936}. Indeed, subsequent experiments performed without current circulating through the wires confirmed the expectation,   implying that the earth's magnetic field had been expelled from the interior of the
superconducting wires. They immediately realized the importance of their accidental discovery and published a short note in October 1933
titled ``Ein neuer Effekt bei Eintritt der Supraleitfahigkeit'',  ``A new effect at the onset of superconductivity'',
where they explicitly stated (guessed?)  that their results implied that the diamagnetic susceptibility became  exactly $-1/4\pi$
in the superconducting state, implying complete expulsion of
the magnetic field. Meissner described in detail the experimental process that led to the discovery in a later publication
in 1936 together with F. Heidenreich \cite{meissner1936}.

 The result was particularly surprising in view of Lippmann's theorem \cite{lipp}, of which Meissner was  very much aware of \cite{meissner1936}, that stated that 
when a normal conductor in the presence of a magnetic field is cooled and becomes a perfect conductor no change in magnetic field
should take place in accordance with Maxwell's equations. There is no indication that neither Meissner nor any other superconductivity
researcher at the time doubted the validity of Lippmann's theorem. In fact, that is presumably the reason for why it took so long, 22 years, from the
discovery of superconductivity in 1911 to the discovery of the  Meissner effect. 
It follows that if  Meissner and Ochsenfeld  had  taken the precaution to shield the earth's magnetic field from their apparatus
with a Helmholtz coil in order to accurately test the
Laue-M\"oglich theoretical prediction without confounding factors, which in fact they had done in other experiments at the time \cite{meissner1936}, the Meissner effect would not have been discovered.

The short Meissner-Ochsenfeld paper \cite{meissner}  contained no figures. A graphical depiction of the Meissner effect was first given by
Meissner and Heidenreich in their 1936 paper \cite{meissner1936}, reproduced here in Fig. 1.

\section{The Keesom experiments}
Immediately after the Meissner effect was discovered it became clear that it had profound implications.
Namely, that the equilibrium state of a superconductor in a magnetic field was independent of its history,
contrary to what had been thought before, and that the superconducting state is an equilibrium thermodynamic state of matter
to which the laws of thermodynamics apply. This was clearly brought to light in the work of 
Gorter \cite{gorter,gorter2} and Gorter and Casimir \cite{gortercasimir}. In particular, it implied that under carefully controlled
conditions the normal-superconductor transition in the presence of a magnetic field is a {\it reversible phase transformation}
with no change in the entropy of the universe, just as any other first order phase transition such as water-ice or water-vapor.

This however raised a puzzling question - not in the minds of theorists that were happy applying the Clausius-Clapeyron formalism
of other first-order phase transformations to the normal-superconductor transition without any qualms, but in the mind of a very thoughtful experimentalist:
Willem Henrik Keesom.

Keesom realized, even if he did not state it explicitly in his papers, that there is a fundamental difference between
any other first order phase transition and the superconducting transition in a magnetic field: in the latter,
 the system carries $momentum$ in one of the phases and does not in the other.

A superconductor in a magnetic field has a surface current that carries mechanical momentum. When it makes a transition to
the normal state the supercurrent stops and its momentum is gone. Where does the momentum go, and how, and what does
it imply?

In a series of  papers between 1934 and 1938 \cite{keesom1,keesom2,keesom3,keesom4,keesom5}  Keesom and coworkers performed careful experiments measuring thermal and
magnetic properties of superconductors in a magnetic field undergoing transitions to and from the superconducting state 
with the purpose of establishing experimentally whether or not the transition was a reversible phase transformation, as
theory \cite{gorter,gorter2,gortercasimir} said it was. In particular, whether any Joule heat was generated  when the system went normal and the supercurrent stopped.
They established quantitatively that indeed to high accuracy no Joule heat is dissipated when the transition occurs
very slowly, implying that the theoretical description as a reversible phase transformation is indeed correct \cite{keesom5,shoenberg}.

Why did Keesom spend so much time and effort in establishing experimentally what seemed obvious from a theoretical point
of view? Presumably because, unlike others, he realized how profoundly counterintuitive it was that a 
supercurrent, with current density much higher than in normal metallic conductors, could stop in the phase transformation to a resistive state
without any dissipation of Joule heat. As he stated  emphatically in one of his papers \cite{keesom2}, 
{\it ``it is essential
that the persistent currents have been annihilated before the material gets
resistance, so that no Joule heat is developed''}. He did not venture to hypothesize how this could happen, 
but deserves the credit for having posed this fundamental question, which surprisingly nobody else paid attention to
until many years later \cite{disapp}. 
 
\section{Thermodynamics of the Meissner effect}

The thermodynamics of the Meissner effect is well understood  \cite{gorter,gorter2,gortercasimir,bcspaper,tinkham,degennes}. I will review it briefly here because it is essential to its understanding.

We consider a long cylindrical superconductor of length $\ell$ and radius $R$, inside a solenoid through which a constant current I circulates,
driven by a power source,
as shown in the lower left inset of Fig. 2, giving rise to a uniform field $H$. The current is given by
\beq
I=\frac{c}{4\pi}\ell H
\eeq
by Ampere's law. We assume that $H=H_c(T_1)$, the thermodynamic critical field  at temperature $T_1$.  When the system is cooled from the normal into the superconducting state, point 1 to point 1' in Fig. 2, both infinitesimally close to $T_1$, the magnetic field
is expelled from its interior. This means that a  current $I$ circulates near the surface of  the superconductor in opposite direction to that of the
solenoid.

       \begin{figure} [t]
 \resizebox{8.5cm}{!}{\includegraphics[width=6cm]{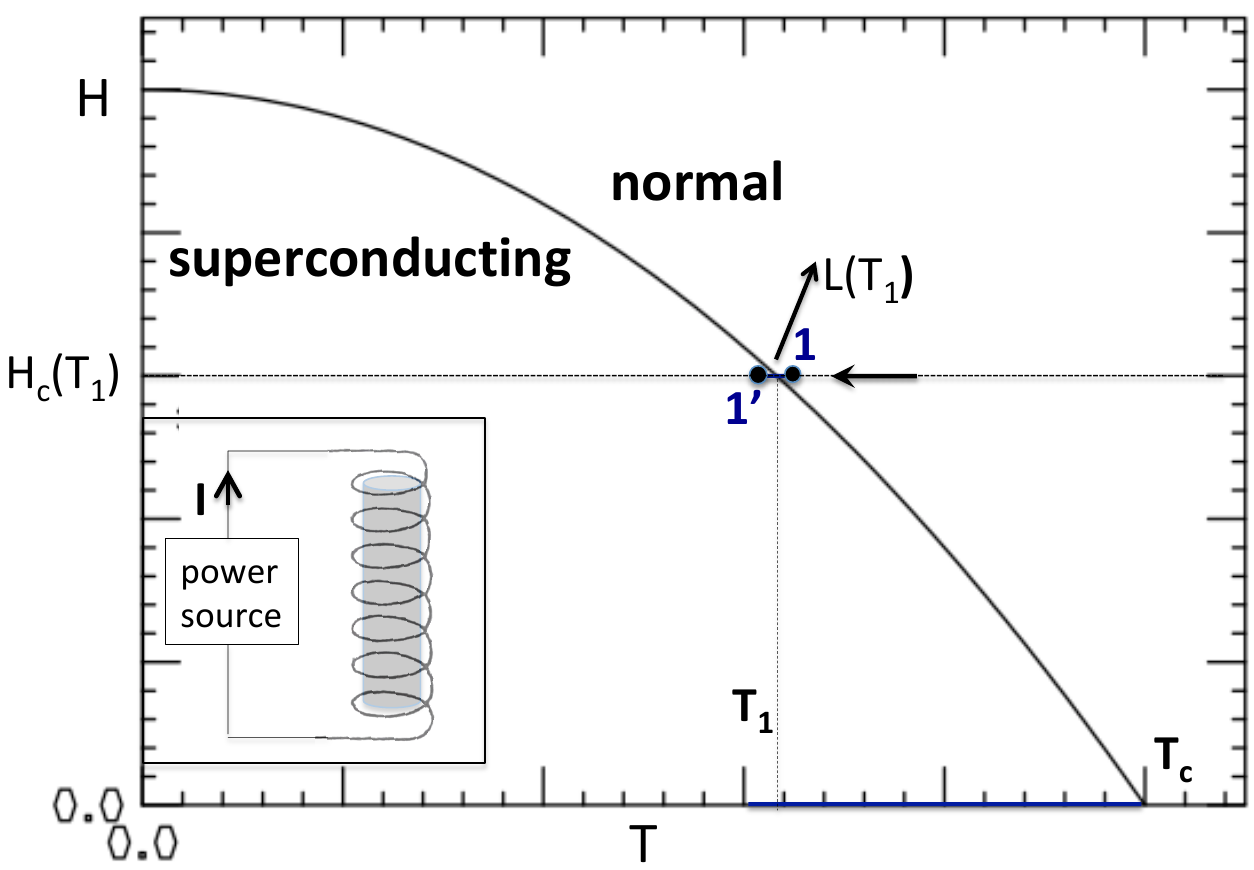}} 
 \caption { Critical field versus temperature for a type I superconductor. The applied external field is $H=H_c(T_1)$.
 The points  {\bf 1} and {\bf 1'} 
  are at temperatures infinitesimally above and below $T_1$. The Meissner effect is the  expulsion of magnetic flux as the system makes
  the transition transition from point {\bf 1} to point {\bf 1'.}  $L(T_1)$ is the latent heat released by the body when it makes the transition from the
  normal to the superconducting state at temperature $T_1$. The lower left inset shows the superconducting sample in a solenoid powered by a battery that
  delivers a constant current I (Eq. (1)).  }
 \label{figure1}
 \end{figure} 

During the process of field expulsion a Faraday counter-emf is generated that opposes the flux expulsion, given by
\beq
\epsilon=-\frac{1}{c}\frac{d\phi}{dt}
\eeq
with $\phi$ the magnetic flux through the superconductor. The Faraday electric field points in the same direction as the current of the solenoid
that keeps the external magnetic field constant, therefore it delivers energy to the power source. The energy per unit time delivered to the power source during the transition  is
\beq
P=\frac{dW}{dt}=\epsilon I= -\frac{1}{c}\frac{d\phi}{dt} (\frac{c}{4\pi}\ell H)
\eeq
and the total energy delivered to the power source as the flux through the sample changes from $\phi(t=0)=\pi R^2 H$ to zero is
\beq
W=\frac{H^2}{4\pi}V
\eeq
with $V=\pi R^2 \ell$ the volume of the sample. From now on we will assume $V=1$ for simplicity. 

Half of that energy, $H^2/(8\pi)$, is the energy of the magnetic field that was in the volume of the sample,
and the other half is the lowering of the energy of the sample in going from the normal to the superconducting state. In addition,
a latent heat $L(T_1)=T_1(S_n(T_1)-S_S(T_1))$ is released by the sample into a heat bath at the same temperature, where
$S_n$ and $S_s$ are the entropies of the normal and superconducting states. 
The change in energy of the sample in going from the normal to the superconducting state is
\beq
U_n-U_s= \frac{H_c(T_1)^2}{8\pi}+T_1(S_n(T_1)-S_s(T_1))
\eeq
and in terms of the free energies ($F=U-TS)$  we have 
\beq
F_n-F_s= \frac{H_c(T_1)^2}{8\pi} .
\eeq
We have ignored the fact that the current in the superconductor circulates not right at the surface but on a surface layer of
thickness $\lambda_L$, the London penetration depth, under the assumption that $\lambda_L<<R$, the radius of the cylinder. 

This thermodynamic understanding implies that the total change in the entropy of the universe is zero in the transition process  \cite{reversible}.
This requires that the transition happens infinitely slowly so that no Joule heat is generated \cite{jouleheat}. 
This requires that the points $1$ and $1'$ in Fig. 2 are infinitely close.

\section{How London electrodynamics explains the Meissner effect }
Let us review briefly the London brothers' F. and H. London description of the Meissner effect \cite{london1,london2,londonbook}, which it is often said explains
the Meissner effect \cite{ai,wiki}, even though it does not. Starting with Newton's equation for the velocity of an electron of mass $m_e$ and charge $e$ under
an electric field $\vec{E}$
\beq
m_e \frac{d \vec{v}_s }{dt}=e\vec{E}
\eeq
gives for the current density $\vec{J}=n_se\vec{v}_s$, with $n_s$ the density of superconducting carriers,
\beq
 \frac{\partial  \vec{J} }{\partial t}=\frac{n_se^2}{m_e}\vec{E}
\eeq
and applying the curl on both sides and using Faraday's law $\vec{\nabla}\times\vec{E}=-(1/c)\partial \vec{B}/{\partial t}$
\beq
 \frac{\partial }{\partial t} \vec{\nabla}\times \vec{J} =-\frac{n_se^2}{m_ec}\frac{\partial \vec{B}}{\partial t}
\eeq
Integrating with respect to time, and assuming as initial conditions no magnetic field and no current yields
\beq
\vec{\nabla}\times \vec{J} =-\frac{n_se^2}{m_ec} \vec{B} =-\frac{c}{4\pi \lambda_L^2} \vec{B} 
\eeq
which is the celebrated London equation, with 
\beq
\frac{1}{\lambda_L^2}\equiv\frac{4\pi n_s e^2}{m_e c^2}
\eeq
giving the London penetration depth $\lambda_L$. From Eq. (10) and Ampere's law it follows that
$\nabla^2\vec{B}=(1/\lambda_L^2)\vec{B}$ which implies that the magnetic field only penetrates a distance $\lambda_L$ into
the superconductor, i.e. is excluded from its interior. 

Of course under the initial conditions appropriate for the Meissner effect, $\vec{J}(\vec{r},t=0)=0$ and
$\vec{B}(\vec{r},t=0)=\vec{B}_0$, time integration of Eq. (9) yields instead of Eq. (10)
\beq
\vec{\nabla}\times \vec{J} =-\frac{c}{4\pi \lambda_L^2} (\vec{B} - \vec{B}_0)
\eeq
with solution $\vec{J}(\vec{r},t)=0$ and
$\vec{B}(\vec{r},t)=\vec{B}_0$ at all times, therefore the initial magnetic field is unchanged, no supercurrent is generated and there is no Meissner effect.

The London brothers $postulated$ \cite{london1,london2} that Eq. (10) describes the state of the superconductor {\it independent of initial conditions}. 
In particular, that the system would reach the state described by Eq. (10) and not by Eq. (12) when cooled
from the normal state  into the superconducting state in the presence of magnetic field $\vec{B}_0$.
The London brothers did not explain what would be the physical reason for the material cooled in the presence of $\vec{B}_0$
to ignore Eq. (12) and instead be governed by Eq. (10). This is still widely misunderstood  today \cite{ai,wiki}.

For the purpose of later discussion, we rewrite London's Eq. (10) in terms of the superfluid velocity as
\beq
\vec{\nabla}\times\vec{v}_s=-\frac{e}{m_ec}\vec{B}
\eeq
or, in terms of the magnetic vector potential $\vec{A}$
\beq
\vec{v}_s=-\frac{e}{m_e c} \vec{A}
\eeq
where we are assuming, following London \cite{londonbook}, that $\vec{A}$ obeys the Coulomb gauge 
$\vec{\nabla}\cdot \vec{A}=0$.

\section{How BCS theory and Ginzburg-Landau theory explain the Meissner effect }

BCS theory \cite{bcspaper,tinkham,degennes} shows that for a system of electrons that interact via a net attractive interaction the ground
state wavefunction is of the form
\beq
|\Psi_{BCS}>=\prod_k(u_k+v_kc_{k\uparrow}^\dagger c_{-k\downarrow}^\dagger)|0> .
\eeq
where $c_{k\sigma}^\dagger$ creates an electron of spin $\sigma$ in the single particle state of crystal momentum
$k$, and $u_k$ and $v_k$ are complex amplitudes determined by minimization of the energy.
BCS showed that  the state Eq. (15) has a lower 
energy than the normal metal Fermi sea. In the presence of a magnetic field $H$ this remains true
up to a critical value of the magnetic field $H_c$ that satisfies
\beq
E_n-E_s=\frac{H_c^2}{8\pi}
\eeq
where $E_n$ is the energy of the normal state with magnetic field in its interior and $E_s$ is the
energy of the superconducting state described by the wavefunction Eq. (15) that does not allow
a magnetic field in the interior. The treatment extends to finite temperatures, leading to the condition
Eq. (6) that determines that if the system is in an applied field $H=H_c(T_1)$ it will have
lower free energy when $T>T_1$ if it is in the normal state with the magnetic field in its interior and it will have lower free energy when $T<T_1$
if it is in the superconducting state with the magnetic field excluded from the interior. 

That is one reason for why it is said that BCS theory predicts and explains the Meissner effect. However the argument is flawed.
It is not a fact that if the system is initially in the normal state at $T>T_1$ and its temperature
is lowered to $T<T_1$ it necessarily will evolve to its lowest free energy state. It will if there is a physical
mechanism for it to do so. If there isn't, it will   stay in the initial state, or it may evolve to another state
that is not the lowest free energy state. The fact that in experiments under field-cooling conditions
the final state invariably has some magnetic field trapped  \cite{mend,shoen,cavities}  indicates that lowest free energy alone
is not a valid criterion to decide where the system will evolve. It is necessary to understand 
the {\it physical processes} that determine how the system goes  from the initial to the final state to understand which final state(s) can
be reached. BCS theory doesn't do that.

Another argument that is used to argue that BCS theory predicts the Meissner effect is the following:
the canonical momentum of a superfluid electron is
\beq
\vec{p}=m_e \vec{v}_s+\frac{e}{c}\vec{A}
\eeq
According to BCS theory the canonical momentum of superfluid electrons in a simply connected sample is $\vec{p}=0$. Setting $\vec{p}=0$ in Eq. (17) yields the London equation Eq. (14), that predicts that no magnetic field
exists in the interior of the superconductor. 

The problem with this argument is of course that it does not
address the question of how the system evolves from the normal state with a magnetic field in the interior
to reach the final state where $\vec{p}=0$. In more detail, in the normal state there is a uniform magnetic field  $\vec{B}_0$ in the
interior of the sample with magnetic vector potential
\beq
\vec{A}(\vec{r},t=0)=\frac{\vec{B}_0\times\vec{r}}{2}
\eeq
and superfluid velocity $\vec{v}_s=0$, so that the canonical momentum is
\beq
\vec{p}(\vec{r},t=0)=\frac{e}{c}\vec{A}(\vec{r},t=0)
\eeq
and how the system evolves to the final state characterized by $\vec{p}(\vec{r},t)=0$ is not described.

 More generally, BCS theory, as well as the associated Bogoliubov-de-Gennes theory \cite{degennes} to treat spatially-dependent situations, 
 and the  phenomenological Ginzburg-Landau theory introduced several years earlier \cite{gl}, imply that the state of the superconductor can be  described by a 
 {\it macroscopic wavefunction} 
 \beq
 \psi(\vec{r})=|\psi(\vec{r})|e^{i\theta(\vec{r})}
 \eeq
 where the amplitude of the wavefunction is the square root of the superfluid density $n_s(\vec{r})$, 
$|\psi(\vec{r})|=n_s(\vec{r})^{1/2}$, and the phase of the superfluid wavefunction is related to the superfluid
velocity through the equation
\beq
 \hbar \vec{\nabla} \theta(\vec{r})=m_e \vec{v}_s(\vec{r})+\frac{e}{c} \vec{A}(\vec{r})
\eeq
which yields the London Eq. (13) upon taking the curl of both sides of Eq. (21). This `proof' applies also to multiply connected samples. But again, the problem is that in the initial stage
of the Meissner effect the system in the normal state is $not$ described by the wavefunction Eq. (20), and
how the system evolves in time to reach the final state described by Eq. (20) that excludes the magnetic field
is not explained by Ginzburg-Landau theory, contrary to what textbooks say \cite{coleman}.

There has been some work extending the Ginzburg-Landau description to time-dependent situations, 
the so-called time-dependent Ginzburg Landau theory (TDGL) \cite{tdgl1,tdgl2}, and this formalism has been used to
describe the normal-superconductor transition in a magnetic field \cite{dorsey, goldenfeld}. In TDGL theory it is assumed that the  time-dependent superconducting
order parameter $\psi(\vec{r},t)$ relaxes exponentially to its equilibrium value in a non-equilibrium situation.
The formalism 
involves a  first order differential equation in time with $real$ coefficients for the time evolution of the order parameter. Hence it describes
$irreversible$ time evolution, and is therefore not relevant to the Meissner effect for type I superconductors,
which is a $reversible$ process under ideal conditions. 
In addition, recent work has shown mathematically  \cite{bcstdgl1,bcstdgl2} that this assumption of TDGL theory
is incorrect and that  within BCS-Bogoliubov-de-Gennes theory the superconducting order parameter will $not$ relax
spontaneously to its equilibrium value.

Another argument claimed to explain the Meissner effect within BCS theory, already given in the original paper  \cite{bcspaper},   starts from the 
system in the BCS state Eq. (15) and calculates the linear response of the system to an applied
magnetic field defined by vector potential $\vec{A}$. The perturbing Hamiltonian is
\beq
H_1=  \frac{ie\hbar}{2mc}     \sum_i  (\vec{\nabla}_i\cdot{A}+\vec{A}\cdot\vec{\nabla}_i) 
\eeq
This perturbation causes the BCS wavefunction $|\Psi_{BCS}>$ to become, to first order in $\vec{A}$
\beq
|\Psi>=|\Psi_{BCS}>-\sum_n\frac{<\Psi_n|H_1|\Psi_{BCS}>}{E_n}|\Psi_n>
\eeq
where  $|\Psi_n>$ are states obtained from the BCS state $|\Psi_{BCS}>$ by exciting 2 quasiparticles, and $E_n$ is the excitation energy. The expectation value of the current operator $\vec{J}_{op}$  with this wave function gives the electric current $\vec{J}$ and the calculation yields \cite{bcspaper,tinkham}:
\beq
\vec{J}=<\Psi|\vec{J}_{op}|\Psi>=-\frac{c}{4\pi}K\vec{A}
\eeq
where $K$ is the `London Kernel' \cite{tinkham}.  
Both $K$ and $\vec{A}$ have wave vector dependence.
which we omit here for simplicity. 
In the long wavelength limit
this calculation yields \cite{bcspaper,tinkham}
$K=1/ \lambda_L^2$ where $\lambda_L$ is the London penetration depth Eq. (11). Eq. (24) is the   London equation
Eq. (10).

In their original paper \cite{bcspaper}, BCS justified the `proof' of the Meissner effect given above with the statement:
{\it ``The electrodynamic properties of our model are determined using a perturbation treatment in which the
first order change in the wave function is used to calculate the current as a functional of the field. For
such properties as the Meissner effect this approach is
quite rigorous since we are interested in the limit as
 $\vec{A}(\vec{r})$ approaches zero.''}
Many researchers in the field immediately questioned the validity of BCS's proof of the Meissner effect.
But, surprisingly, for the wrong reason. They focused on the fact that Eq. (24) used by BCS is not gauge
invariant, and argued that the proof has to be valid in a gauge invariant way. After considerable
mathematical elaborations, several workers independently proved that the non-gauge invariant 
BCS proof can be reformulated in a gauge-invariant way \cite{schafroth,andersongauge,rick} and remains valid. This casted in stone the firm belief of the
scientific  community, ever since then to the present, that BCS theory has fully explained the Meissner effect.

However, the real weakness of the BCS justification for their proof of the Meissner effect is immediately 
apparent upon closer examination of their statement reproduced above. Their approach is not
{\it ``quite rigorous''}, and the fact that they are 
{\it ``interested in the limit as
 $\vec{A}(\vec{r})$ approaches zero''}  is   irrelevant. The response of the system to 
a magnetic field, no matter how small it is, should be calculated with the system starting in its
initial state, namely the normal state, not the unperturbed BCS state. The system cannot 
reach the BCS state while the magnetic field is present, no matter how small it is, simply because
the BCS state requires macroscopic phase coherence and any infinitesimal magnetic field will
destroy the phase coherence. Therefore, the BCS proof of the Meissner effect \cite{bcspaper} proves nothing
about the Meissner effect.

Finally, it is  often said  that
the Meissner effect is the same as the Higgs mechanism in elementary particle physics and is explained by the fact that  massless photons become massive when a metal goes
superconducting \cite{andersonhiggs,coleman,sethna,weinberg,weinberg2}. The argument is based on Ginzburg-Landau theory and also assumes that the system has a well-defined phase at
the outset, hence doesn't address the question of how the system reaches that state starting from the normal state. Therefore
this also says nothing about how and why the Meissner effect occurs in materials.
 
 The issue of reversibility, discussed in Sect. IV,  is key to the Meissner effect. Under ideal conditions, the transition is thermodynamically reversible with no
change in the entropy of the universe. This was well understood immediately after the Meissner effect was discovered
 \cite{gorter,gorter2,gortercasimir, shoenberg}, and it is also 
implicit in the BCS description of the phenomenon. Therefore, any valid description of the Meissner effect needs to be able
to describe a $reversible$ process. This seems to have been forgotten in more recent times \cite{dorsey, goldenfeld,goldenfeld2},
but will be central in the considerations that follow.

\section{Puzzles with momentum in  the Meissner effect and its reverse}

In the above recount of arguments underlying the belief that the conventional theory of superconductivity 
explains the Meissner effect there 
was no mention of momentum. However, momentum is clearly involved in the Meissner effect and needs to be understood.
The Meissner current that shields the external magnetic field carries mechanical momentum. The momentum density
associated with electric current density $\vec{J}(\vec{r})$ is
$\vec{\mathcal{P}}(\vec{r})=(m_e/e)\vec{J}(\vec{r})$, with $m_e$ the $bare$ electron mass and $e$ the electron charge. For the cylinder of radius $R$ and length $\ell$ considered in Sect. V, the total angular momentum carried
by the supercurrent in an external field $H$ is \cite{momentum}
\beq
\vec{L}_e=-\frac{m_e c}{2e}\ell R^2 H \hat{z}  .
\eeq
This is a macroscopic angular momentum    that can be (and has been)  measured experimentally \cite{kikoin,gyro,doll}.
For a numerical example, for $R=1cm$, $\ell=5cm$ and $H=200G$, $L_e=2.84 mg \cdot mm^2/s$. How this momentum originates and is conserved needs to be explained.

Fig. 3 shows in panel {\bf a}  a superconductor to which a magnetic field is applied,
and the resulting state is shown in panel {\bf b}. The Faraday electric field generated in the process, pointing clockwise
as seen from the top, creates
the surface current that prevents the field from penetrating. This is given quantitatively for free electrons by the 
London equation Eq. (10)  reviewed in Sect VI. For electrons in a periodic potential the result is still Eq. (10) but with the
electron bare mass in the London penetration depth Eq. (11)  replaced by the band effective mass \cite{momentum}. The current flows in clockwise direction and the electrons move in
counterclockwise direction, as dictated by the Faraday field, carrying electronic angular momentum pointing up (parallel to the applied field). At the same time the Faraday electric field imparts
clockwise  momentum to the positively charged ions, and it is easily shown assuming charge neutrality that the body acquires
clockwise momentum that exactly cancels the electronic angular momentum   \cite{momentum}, as required.
So this is all consistent and easy to understand.

        \begin{figure} [t]
 \resizebox{8.5cm}{!}{\includegraphics[width=6cm]{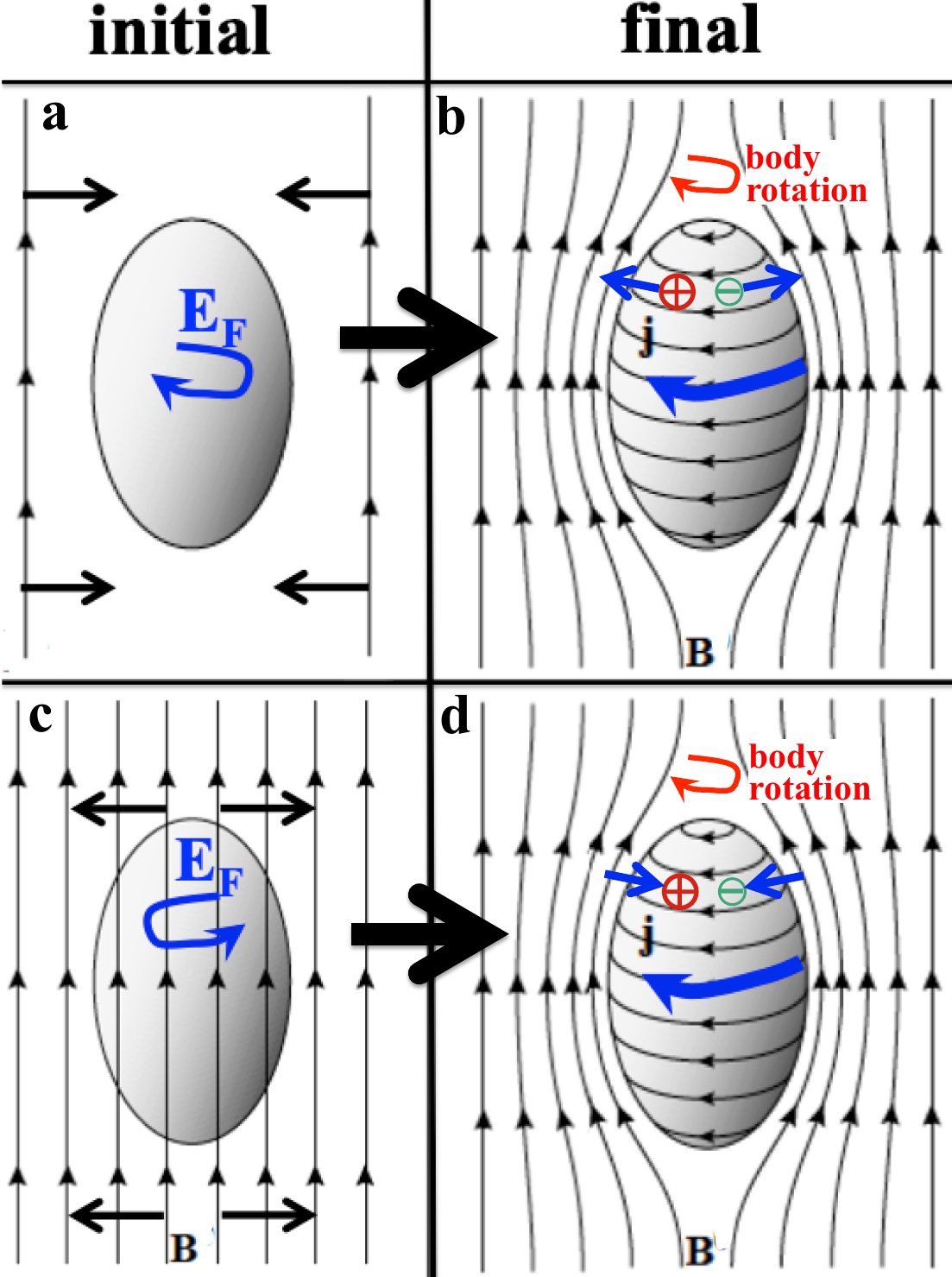}} 
 \caption { Panels {\bf a} and   {\bf b} show application of a magnetic field to a simply connected body that is
 already in the superconducting state.  The Faraday electric field $E_F$ points clockwise and gives rise to the clockwise current
 {\bf j} that prevents the magnetic field from penetrating, and to clockwise rotation of the body. 
 Panels {\bf c} and   {\bf d} show the Meissner effect. The Faraday electric field points   in counterclockwise direction,
 yet the momentum acquired by the electrons and the body as a whole are the same as in panel {\bf b}. 
 The blue arrows on the electron and the ion on panels  {\bf b} and  {\bf d} indicate the
 direction of the force exerted by the Faraday field. . }
 \label{figure1}
 \end{figure}

Let us now consider  the Meissner effect, shown in panels {\bf c}  and {\bf d}  of Fig. 3. The system is cooled into
the superconducting state with magnetic field lines in the interior, which have to move out to reach the final state,
so the Faraday electric field points now in opposite direction during the process, i.e. counterclockwise. 
The Faraday electric field pushes electrons to move clockwise and ions to move counterclockwise, as shown
in Fig. 3 {\bf d}.
Yet the final state that is reached is identical to the one shown in panel {\bf b}, i.e. electrons moving counterclockwise
and the body rotating clockwise.

So it has to be explained: (1) How did the  electrons in Fig. 3 {\bf d}  acquire momentum in direction opposite to that imparted by the Faraday
electric field, and (2) how do the ions and hence the body as a whole acquire momentum in direction opposite to that imparted by the Faraday electric field? These questions need to be answered. 

Similar puzzles occur in the inverse transition, from superconducting to normal, shown in Fig. 4. The Faraday field points
clockwise, attempting to keep the current going, and pushes the body to rotate clockwise. Still, in the final state (left panel of Fig. 4)
 the body rotates counterclockwise, since it inherited the momentum of the electrons in the supercurrent, and the supercurrent has stopped. It has to be explained: (1) How did  the supercurrent stop, 
against the Faraday field that pushed it to keep going, and (2) how does the body acquire momentum
 opposite to that imparted by the Faraday field on the ions.

        \begin{figure} [t]
 \resizebox{8.5cm}{!}{\includegraphics[width=6cm]{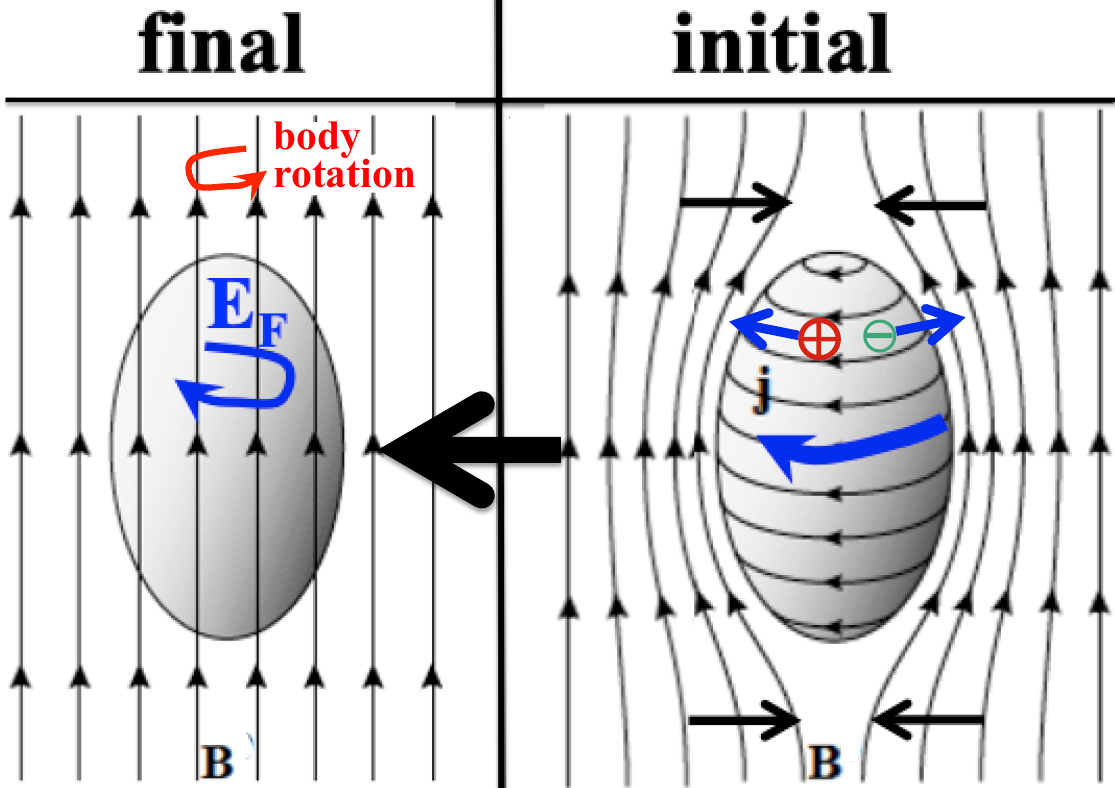}} 
 \caption {Inverse of the Meissner transition, right panel to left panel. System in a magnetic field goes from superconducting to normal.
 The Faraday field points clockwise, attempting to keep the clockwise current {\bf j} going and pushing the body to
 rotate in clockwise direction.  In the final state (left panel) the current has stopped and the body rotates 
 counterclockwise. }
 \label{figure1}
 \end{figure}

It should be emphasized that any processes involving momentum transfers between electrons and the body
as a whole need to be $reversible$ processes, consistent with the fact that the Meissner effect is
thermodynamically reversible under ideal conditions, as discussed in Sect. V.  In this connection it is important to consider that 
it is experimentally found that for a material to undergo a complete Meissner effect it has to be extremely pure  \cite{mend,shoen,cavities} . Any imperfections will lead to the diamagnetic moment
under field cooling to be smaller than that under zero field cooling conditions, implying that some flux was not expelled but remained
trapped in the interior of the material. This by itself strongly suggests that the mechanisms
responsible for  momentum transfer processes between electrons and ions under ideal conditions in the normal-superconductor
transition in a magnetic field are not related to scattering mechanisms that give rise to finite resistance in the normal state
such as impurities and imperfections that break the lattice periodicity, since those
scattering processes are irreversible.

There  has been  one attempt to describe the superconductor to normal transition in a magnetic field
within the conventional framework using   time-dependent  Ginzburg-Landau theory \cite{tdgl1,tdgl2} that took into account  the momentum transfer between electrons and ions \cite{eilen}.
A term in the current density in that formalism describes the current carried by  normal electrons,
stemming from the momentum transferred to the normal electron fluid when
Cooper pairs carrying center of mass momentum unbind and the superfluid electron density decreases.
The author   states \cite{eilen}  that  {\it ``this momentum then decays with the transport relaxation time $\tau$''}. However, such decay would necessarily lead
to Joule heat dissipation and hence irreversibility, therefore this approach cannot be correct. More generally,  any approach that assumes that the momentum of the Cooper
pair is transferred to normal quasiparticles when the system goes from superconducting to normal cannot be correct since in normal metallic transport decay of electric current is necessarily associated with Joule heat,
thermodynamic irreversibility  and even electromigration for high current densities. 
Rather,  as correctly anticipated by Keesom early on \cite{keesom2}, {\it ``“it is essential that the persistent currents have been annihilated before the material gets resistance''}.

 \section{The puzzles in more detail}
 We have seen in the last section that the momentum transfer between electrons and ions in the Meissner
 effect, as well as in the reverse process, the superconductor to normal transition, is opposite to what is dictated by Faraday's law and its mechanism needs to be explained. In particular, in the superconductor to normal transition
 the initial momentum of the supercurrent, Eq. (25), needs to be transferred to the body as a whole without dissipation, and the initial kinetic energy of the
 supercurrent needs to go somewhere without dissipation. The puzzle of how the supercurrent stops without dissipation was
  first identified by Keesom, as discussed in Sect. IV.
 In fact, as we will show in this section, the total momentum transfer between electrons and ions that needs to occur and needs to
 be explained is many
 orders of magnitude larger than $\vec{L}_e$ given by Eq. (25), and the total conversion of kinetic energy of the supercurrent to another energy
(that is $not$ Joule heat) that has to occur is many orders of magnitude larger than the initial kinetic energy of the supercurrent.
 
           \begin{figure*} [t]
 \resizebox{14.5cm}{!}{\includegraphics[width=6cm]{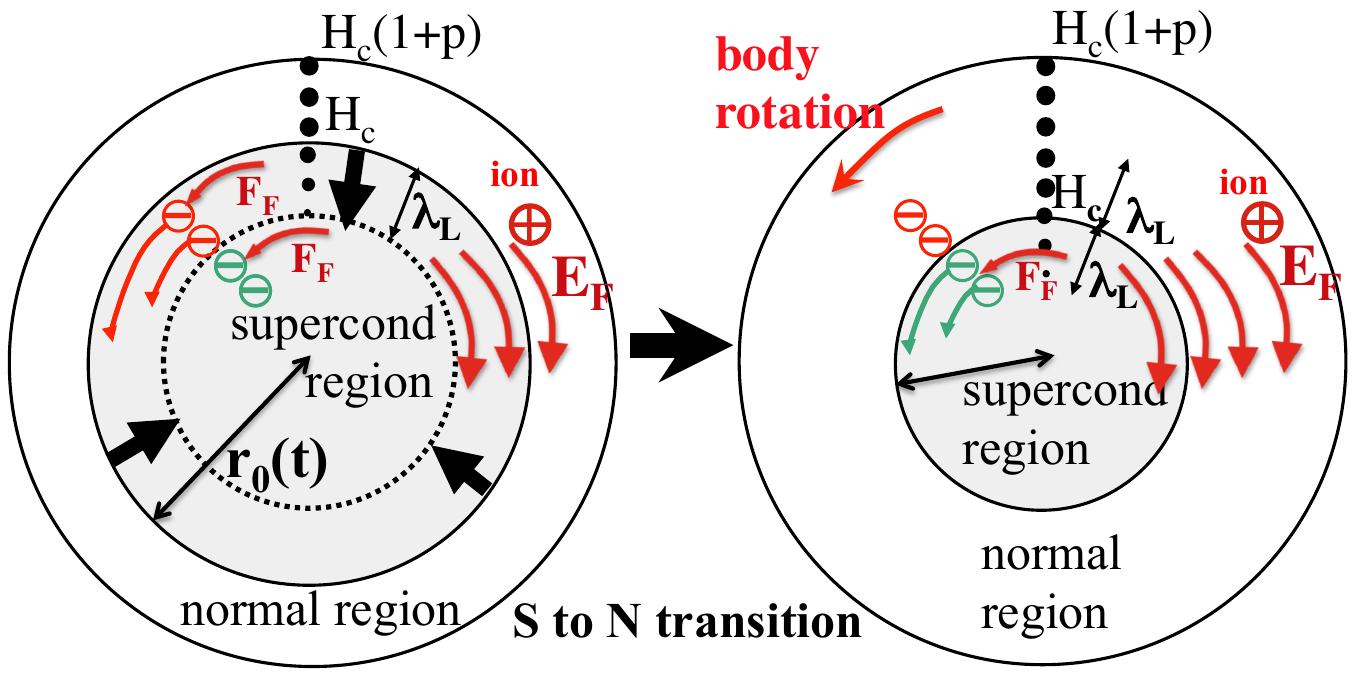}} 
 \caption {Superconductor to normal transition in a cylinder as seen from the top. The magnetic field points out of the picture. On the left panel the phase boundary is at  radius $r_0(t)$, and supercurrent circulates (approximately) in the region
 $r_0(t)-\lambda_L< r< r_0(t)$. On the right panel the phase boundary has moved to $r_0(t)-\lambda_L$,
 the supercurrent that was in the region $r_0(t)-\lambda_L< r< r_0(t)$ has stopped and now there is supercurrent in the region
 $r_0(t)-2 \lambda_L<r<r_0(t)-\lambda_L$. $E_F$ is the Faraday electric field, and $F_F$ denotes the  force that $E_F$ exerts on electrons,
 $F_F=eE_F$.
 The puzzle is, how did the red electrons on the right panel of the figure lose their kinetic energy and angular momentum in going from
 the superconducting region to the normal region as the phase boundary crossed them.
 And, how does the body acquire  counterclockwise momentum
 when the Faraday field pushes the ions in clockwise direction.}
 \label{figure1}
 \end{figure*} 
 
 Let us consider the time evolution of the phase boundary between superconductor and normal phases in the S to N transition in a
 cylindrical geometry for simplicity. The kinetics of the process for the superconductor to normal transition was first discussed by Pippard in a seminal paper in
 1950 \cite{pippardpaper}. Pippard showed that what determines the rate of the transition can be simply calculated from
 electromagnetic considerations. Upon application of a magnetic field $H_c(1+p)$ slightly larger than $H_c$  to the boundary, the magnetic field
 starts to penetrate at a speed such that the resulting Faraday electric field induces an opposite magnetic field in the growing normal region that reduces the
 value of the magnetic field at the boundary to exactly $H_c$. That speed is inversely proportional to the electrical conductivity of the normal region. 
Pippard  deduced the approximate relation for the total transition time to be $\mathcal{T}=\pi \sigma R^2/p$ with $R$ the cylinder radius. The total amount of Joule heat dissipated is
 inversely proportional to $\mathcal{T}$ and the transition becomes reversible when it proceeds infinitely slowly, which will
 happen if $p\rightarrow 0$.
 
 Let us analyze quantitatively the energy and momentum transfers in this process, shown in Fig. 5. We assume the phase boundary between normal and superconducting phases remains
 cylindrical. The supercurrent circulates within a distance approximately $\lambda_L$ from the phase boundary of radius $r_0(t)$ where the applied magnetic field
 penetrates. The magnetic field is given by
 \beq
 B(r,t)=H_ce^{(r-r_0(t))/\lambda_L}
 \eeq
 for $r\le r_0(t)$, and the velocity of  the electrons carrying the supercurrent is
 \beq
 v(r,t)=\frac{e\lambda_L}{m_e c}H_c e^{(r-r_0(t))/\lambda_L} .
 \eeq
 With $n_s$ superconducting electrons per unit volume the kinetic energy of the supercurrent is
 \beq
 K_e=2\pi n_s \ell \int_0^{r_0} dr r \frac{1}{2}m_e v(r,t)^2=
\frac{H_c^2}{8\pi} \pi \ell r_0 \lambda_L .
 \eeq
 At the beginning of the process, with $r_0=R$, the initial kinetic energy of the supercurrent is
 \beq
 K_e=\frac{H_c^2}{8\pi} \pi \ell R \lambda_L =\frac{H_c^2}{8\pi} V \frac{\lambda_L}{R}
 \eeq
 with $V=\pi R^2 \ell$ the volume of the superconductor.
When the phase boundary moves inward a distance $dr_0$ from $r_0$, the superfluid electrons in the range
 $r_0-dr_0<r<r_0$ enter the normal region and stop their motion. More precisely, the center of mass momentum of the Cooper pair
 becomes zero as the pair unbinds. The change in kinetic energy
 when the superelectrons in the volume $2\pi r_0 dr_0 \ell$
 go normal is
 \beq
 dK_e = \frac{H_c^2}{8\pi}(2\pi r_0 \ell)dr_0
 \eeq
 and the total change in kinetic energy as the phase boundary moves from $r_0=R$ to $r_0=0$ is
 \beq
 K_e^{tot}=\int _0^R dK_e=\frac{H_c^2}{8\pi} V .
 \eeq
 Thus,
 \beq
 K_e^{tot}=\frac{R}{\lambda_L} K_e
 \eeq
 with $K_e$ the initial energy of the supercurrent.
  The quantity Eq. (31) is the difference in free energies between normal and superconducting states, Eq. (6).

What Eqs. (31) and (32) say can be understood with the help of Fig. 5. As the phase boundary moves inward, a clockwise Faraday electric field
$E_F$ is generated that wants to preserve the supercurrent. The field exerts a counterclockwise force $F_F$ on superelectrons in the interior of the superconducting
region that thereby acquire kinetic energy density $H_c^2/8\pi$ right when the phase boundary that is moving inward reaches them.
At that point Cooper pairs dissociate, the electrons become normal with zero net momentum, and the kinetic energy of the supercurrent is used up
in the dissociation of Cooper pairs, i.e. is the difference in free energies between normal and superconducting states given by Eq. (6).
The total kinetic energy that is converted into free energy in the entire process, Eq. (32),  is a factor $R/\lambda_L$ larger than the initial kinetic energy of the
supercurrent, e.g. $250,000$ times larger for a typical example $R=1cm$, $\lambda_L=400 \AA$. None of that kinetic energy gets
dissipated as Joule heat.

Let us verify that the Faraday electric field is responsible for imparting the kinetic energy to the supercurrent. From Faraday's law and Eq. (26) for the
magnetic field it follows that for $r\le r_0$ 
\beq
E_F(r,t)=\frac{\lambda_L}{c}\frac{\partial}{\partial t} H_c e^{(r-r_0(t))/\lambda_L}=\frac{\dot{r}_0}{c}H_c e^{(r-r_0(t))/\lambda_L}
\eeq
and the change in velocity for superfluid carriers is given by
\beq
\frac{\partial }{\partial t} v(r,t)       =eE_F(r,t)=\frac{e\lambda_L}{c} \frac{\partial}{\partial t} H_c e^{(r-r_0(t))/\lambda_L}
\eeq
which yields Eq. (27) upon time integration.

We can similarly understand what happens with the angular momentum transfer. The initial angular momentum of the supercurrent is $L_e$, Eq. (25).
As the phase boundary moves inward the Faraday field imparts angular momentum to the supercarriers in the same direction. 
The angular momentum of the supercurrent when the phase boundary is at radius $r_0$ is
\beq
 L_e(r_0)=2\pi n_s \ell \int_0^{r_0} dr r (m_e v(r,t) r)=\frac{m_ec}{2\pi e} (\pi r_0^2 \ell) H_c.
 \eeq
 When the phase boundary moves inward by $dr_0$ the supercurrent in that region stops. Its change in angular momentum is
 \beq
 dL_e=\frac{m_ec}{2\pi e} (\pi r_0^2 \ell)H_c\frac{dr_0}{\lambda_L}
 \eeq
 so that the total change in angular momentum of electrons when the system  goes from the superconducting to the normal state is
 \beq
 L_e^{tot}=\int_0^R dL_e= \frac{m_e c}{2\pi e} (\pi R^2 \ell)H_c \frac{R}{3\lambda_L}
 \eeq
 hence
 \beq
  L_e^{tot}= \frac{R}{3\lambda_L} L_e .
\eeq
with $L_e=L_e(R)$ given by Eq. (25).
For the example given above, $L_e^{tot}=83,333 L_e$.

The angular momentum of electrons in the supercurrent is counterclockwise. So an amount of electronic angular momentum which is 5 orders of magnitude larger than the initial angular momentum of the supercurrent, almost all of it created by action of the Faraday electric field on the current carriers,
disappears from the electronic system in the process of the transition from the superconducting to the normal state.
Of course it cannot disappear because angular momentum is conserved, hence must be transferred to the body as a whole.
At the same time, the Faraday electric field imparts $83,332L_e$ clockwise angular momentum to the body as it acts on the ions in the superconducting
region close to the phase boundary. The net result is that the body inherits exactly $L_e$ of angular momentum from the electrons, the initial angular momentum
of the supercurrent.

             \begin{figure*} [t]
 \resizebox{14.5cm}{!}{\includegraphics[width=6cm]{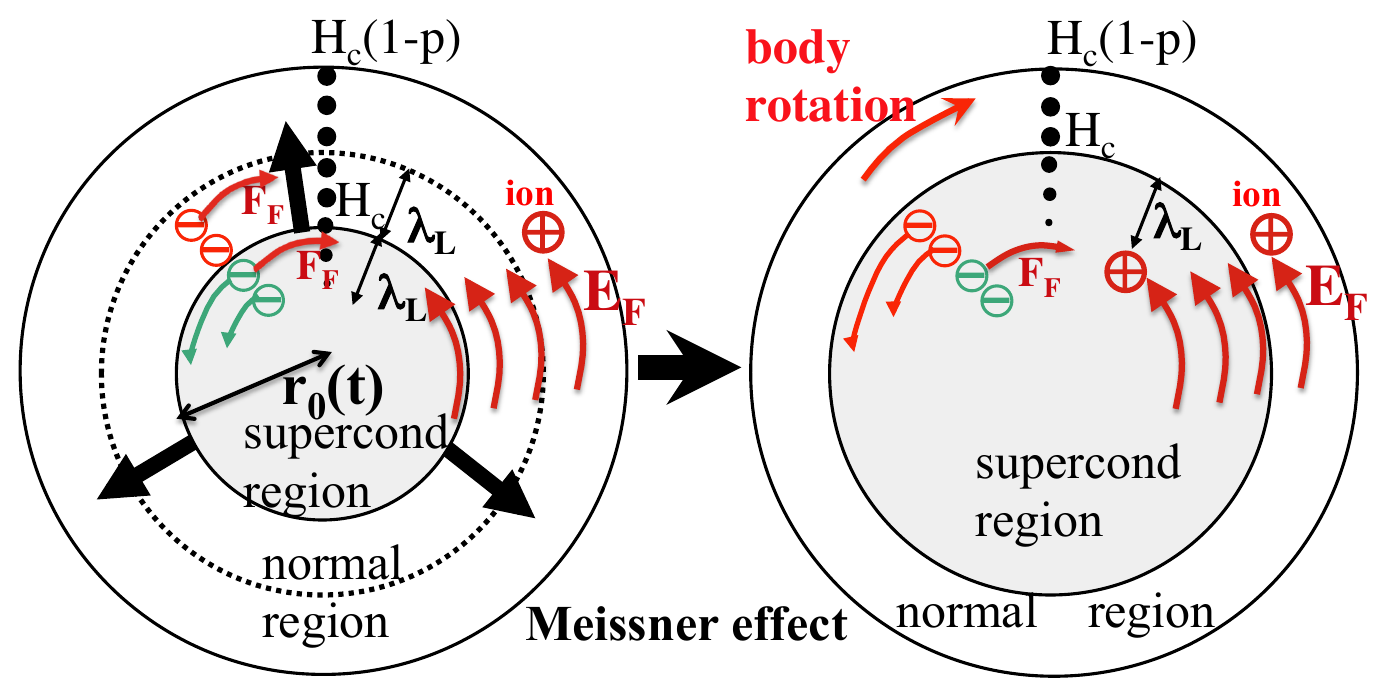}} 
 \caption {Normal to  superconductor transition in a cylinder as seen from the top. The magnetic field points out of the picture. On the left panel the phase boundary is at  radius $r_0(t)$, and supercurrent circulates (approximately) in the region
 $r_0(t)-\lambda_L< r< r_0(t)$. On the right panel the phase boundary has moved to $r_0(t)+\lambda_L$,
 supercurrent   in the region $r_0(t)< r< r_0(t)+\lambda_L$ was created.   $E_F$ is the Faraday electric field, that slows down electrons in the superconducting
 region as the phase boundary moves further outward. 
 The puzzle is, how did the red electrons on the right panel of the figure acquire their   kinetic energy and angular momentum in going from
 the normal to the superconducting region  as the phase boundary crossed them. And, how does the body acquire  clockwise momentum
 when the Faraday field pushes the ions in counterclockwise direction.}
 \label{figure1}
 \end{figure*}

The mechanism by which the body acquired $83,332L_e$ clockwise angular momentum is clear, it is the momentum imparted to it by the Faraday
electric field. The mechanism by which the electrons in the supercurrent when they entered the normal region stopped and  transferred
$83,333L_e$ counterclockwise angular momentum to the ions needs to be explained. Once again, it cannot entail 
normal electrons resulting from dissociation of Cooper pairs inheriting the center of mass momentum of the Cooper pairs and then   transfering it to the body through scattering, because that would generate Joule heat. 
Joule heat is not allowed for two reason: one is that it would make the process irreversible, and in addition it would violate conservation of energy
since all the kinetic energy of the supercurrent was accounted for in the free energy difference between superconducting and normal states, as
discussed above.

For the Meissner effect, where the processes just discussed run in reverse, the same puzzles arise. Fig. 6 shows a step in the normal to superconducting transition. When the phase boundary moves outward a distance $dr_0$ from $r_0$, the normal electrons in the range
 $r_0<r<r_0+dr_0$ enter the superconducting region and suddenly acquire momentum. More precisely the Cooper pair as it forms acquires center of mass momentum. That center of mass momentum corresponds to each electron in the 
 Cooper pair acquiring an azimuthal velocity
 \beq
 v(r_0)=\frac{e\lambda_L}{m_e c} H_c 
 \eeq
 and kinetic energy,  using Eq. (11) 
 \beq
 k_e=\frac{1}{2} m_e v(r_0)^2=\frac{e^2 \lambda_L^2}{2m_e c^2}H_c^2=\frac{H_c^2}{8\pi n_s}
 \eeq
as they enter the superconducting region. The change in kinetic energy
 when the normal electrons in the volume $2\pi r_0 dr_0 \ell$
condense into Cooper pairs is
 \beq
 dK_e =k_e n_s(2\pi r_0 \ell)dr_0=  \frac{H_c^2}{8\pi}(2\pi r_0 \ell)dr_0 ,
 \eeq
and the total change in kinetic energy as the phase boundary moves from $r_0=0$ to $r_0=R$ is
 \beq
 K_e^{tot}=\int _0^R dK_e=\frac{H_c^2}{8\pi} V .
 \eeq
   The quantity Eq. (41) is the difference in free energies between normal and superconducting states, Eq. (6).

What Eqs. (39-42) say can be understood with the help of Fig. 6. As the phase boundary moves outward, normal electrons condensing into the superconducting state suddenly acquire kinetic energy, so their condensation energy is converted into kinetic energy in the condensation process. 
A counterclockwise Faraday electric field
$E_F$ exists as the magnetic field lines are moving out. The Faraday field exerts a clockwise force $F_F$ on superelectrons   that thereby lose their kinetic energy Eq. (40) as the phase boundary moves further away from them. The same Eqs. 
(33) and (34) explain how the Faraday electric field slows down the electrons so that they lose their 
entire kinetic energy as they become part of the deep interior of the superconducting phase where no current nor magnetic field exists.
The total free energy that is converted into kinetic energy in the entire process is a factor 
$R/\lambda_L$ larger than the final kinetic energy of the supercurrent Eq. (29).

We can similarly understand what happens with the angular momentum transfer. 
The angular momentum of the supercurrent when the phase boundary is at radius $r_0$ is
\beq
 L_e=2\pi n_s \ell \int_0^{r_0} dr r (m_e v(r,t) r)=\frac{m_ec}{2\pi e} (\pi r_0^2 \ell) H_c.
 \eeq
 When the phase boundary moves outward by $dr_0$ the electrons in the region 
 $r_0<r<r_0+dr_0$ acquire angular momentum through an unknown process. The change in angular momentum is
 \beq
 dL_e=\frac{m_ec}{2\pi e} (\pi r_0^2 \ell)H_c\frac{dr_0}{\lambda_L}
 \eeq
 so that the total angular momentum acquired by electrons in the Meissner process is 
 \beq
 L_e^{tot}=\int_0^R dL_e= \frac{m_e c}{2\pi e} (\pi R^2 \ell)H_c \frac{R}{3\lambda_L}
 \eeq
 hence
 \beq
  L_e^{tot}= \frac{R}{3\lambda_L} L_e 
\eeq
with $L_e$ the angular momentum of the supercurrent at the end of the process, Eq. (25).
For the example given above, $L_e^{tot}=83,333 L_e$.
That angular momentum has to be transferred to the body as a whole by an unknown mechanism
that cannot involve scattering that would lead to dissipation and irreversibility.
The Faraday electric field plays the role of imparting angular momentum to electrons and ions
in the amount $L_e^{tot}=83,332 L_e$ in opposite directions, so that at the end
electrons and ions have angular momentum $L_e$ in opposite directions, ensuring momentum conservation.

 \section{Electromagnetic energy flow}
According to Poynting's theorem, 
\beq
\oint \vec{S}\cdot d\vec{a}=-\frac{\partial}{\partial t} \int d^3r u - \int d^3r \vec{E}\cdot \vec{J} .
\eeq
with
\bmath
\beq
\vec{S}=\frac{c}{4\pi} \vec{E}\times\vec{B}
\eeq
\beq
u=\frac{1}{8\pi} (E^2+B^2).   
\eeq
\emath
Let us apply this theorem to the surface of a small cylindrical region  of radius $r_0$ and height $h$ in the interior of the metal 
going from normal to superconducting in the presence of a magnetic field at temperature $T_1$. 
The vector $d\vec{a}$ is the normal to the surface pointing outward, and the left-hand-side of Eq. (47) is the outflow of electromagnetic energy.
The first term on the right-hand-side of Eq. (47) is the decrease in electromagnetic energy inside the volume enclosed by the surface,
and the second term is minus the work done by the electric  field on charges inside the volume: if the electric field does negative work on the charges,
the outflow of electromagnetic energy increases.
The electric field on the lateral surface that is induced as the magnetic  flux $\phi$ through the region changes is
\beq
\vec{E}(r_0)=-\frac{1}{2\pi r_0 c}\frac{\partial \phi}{\partial t}\hat{\theta}
\eeq
pointing in counterclockwise direction (as seen from the top), so the Pointing vector is parallel to the outward normal to the surface $d\vec{a}$.
Assuming  that the magnetic field at the surface of the region stays constant at $H=H_c(T_1)$  as the magnetic field  inside the region is expelled, the left side of Eq. (47) is
\beq
 \oint \vec{S}\cdot d\vec{a}=-\frac{1}{4\pi} h H \frac{\partial \phi}{\partial t}=-\frac{1}{4\pi} h \pi r_0^2  H \frac{\partial B}{\partial t}
\eeq
The first term on the right-hand-side of Eq. (47) is, assuming the magnetic field is uniform inside the region
\beq
-\frac{\partial}{\partial t} \int d^3r u=-\frac{1}{8\pi} h \pi r_0^2 \frac{\partial B^2}{\partial t} 
\eeq
and Eq. (47) yields, with $V_r=\pi r_0^2 h$ the volume of the region
\beq
-\frac{1}{4\pi}   H \frac{1}{\pi r_0^2}  \frac{\partial \phi}{\partial t}= -\frac{1}{8\pi} \frac{\partial B^2}{\partial t} -\frac{1}{ V_r}\int d^3r \vec{E}\cdot \vec{J} .
\eeq
Integrating over time and using that $B(t=0)=H$, $B(t=\infty)=0$ in the interior of the  region  yields
\beq
\frac{1}{4\pi}   H^2 =\frac{1}{8\pi} H^2  - \frac{1}{V_r}\int_0^\infty dt\int d^3r \vec{E}\cdot \vec{J} .
\eeq
Eq. (53) shows that the magnetic field cannot be expelled from the interior of the region unless the electric field in the interior of the  region performs
negative work on charges inside the region    in the amount $-H^2/(8\pi) V_r$.  
Eq. (53) will be satisfied as the material inside the region  goes from the normal to the superconducting state because a supercurrent flowing clockwise gets generated and the Faraday field $E$ pointing counterclockwise
performs negative work on the supercurrent  amounting to precisely $-H^2/(8\pi)V_r$ \cite{scstops}, rendering Eq. (53) an equality.

The above conclusion was reached assuming  that the magnetic field at the boundary of the region remains unchanged as the field is expelled from its  interior.
If instead the magnetic field at the boundary of the region was  to go to zero as the magnetic field is expelled, Eq. (47) would be satisfied with no contribution from
its second right-hand term, i.e. no current in the interior of the region. This would happen if, for example, the Meissner current expelling the magnetic field
would develop at the boundary of the cylinder, so that the magnetic field would decrease uniformly inside the cylinder including the region under
consideration
and its surface. This is however impossible even within the conventional theory, since in order for the current at the boundary of the cylinder to overcome the 
Faraday counter-emf requires condensation energy from the  interior, which is not available while any magnetic field is still in the interior. 
More generally we note that  Eq. (47) implies  that for the magnetic field to go from finite to zero at any point inside the volume requires that
 at some point during the process there was electric current at that point, whose carriers gained kinetic energy through 
 the condensation process and  subsequently delivered that energy to the electromagnetic field through the 
 second term
 on the right-hand side of Eq. (47).

This analysis  indicates that magnetic field expulsion is  a $local$ process.
Magnetic field will get excluded locally from the region that becomes superconducting as it becomes superconducting.
Thus, the transition must proceed through nucleation and growth of the lower temperature phase, just as, for example, 
the water-ice transition.  However, unlike the water-ice or any other first order phase transition, here there is $momentum$ involved as well as {\it kinetic energy}. A nucleating superconducting region acquires a current circulating around its boundary
carrying momentum and kinetic energy 
that did not exist in the normal phase. This raises questions about conservation laws \cite{momentum}  that need to be answered
if one is to claim that the process is understood \cite{bcs50}, questions that do not arise in other
first order phase transitions where only potential energy is involved.
 
\section{Maxwell pressure and Meissner pressure}
The Maxwell stress tensor in the absence of electric field is
\beq
T_{ij}=\frac{1}{8\pi}(H_iH_j-\delta_{ij} H^2)
\eeq
and in the radial direction it is given by
\beq 
T_{rr}=-\frac{1}{8 \pi}H_c^2\equiv -P_{Ma}
\eeq
This   ``Maxwell pressure'' $P_{Ma}$ resists the magnetic field expulsion at every point
inside the volume.  At a point where the condensation starts, a ``Meissner pressure''  \cite{londonbook} $P_{Me}$
needs to be generated pointing outward as the   superconducting phase nucleates and
subsequently grows, with the Meissner pressure slightly overpowering    the Maxwell pressure (if it happens reversibly). 
 The term ``Meissner pressure'' was coined by London \cite{londonbook}, who realized the necessity for its
existence but did not provide a physical explanation for its origin.

The Maxwell pressure manifests itself by the radial Lorentz force exerted by the magnetic field
on the clockwise circulating current around the boundary of a domain that nullifies the magnetic field
in its interior.
This inward force $F_r$ on an electron \cite{ondyn,koizumi}  is   given by
\beq
\vec{F}_r=\frac{e}{c}\vec{v}\times\vec{B}
\eeq
with $\vec{v}=v_\theta(r)\hat{\theta}$ the azimuthal velocity of the electrons in the supercurrent near the boundary of the domain,
flowing in counterclockwise direction. $\vec{F}_r$ represents the transfer of momentum from the electromagnetic field
described by $T_{rr}$
to the electrons.

For the domain to grow rather than shrink, the system as it condenses needs to do work against
this radial   force  that acts  on  the electrons in the supercurrent near the surface of the domain. For a circular
domain of radius $r_0$, the velocity of the electrons is given by Eq. (27), 
and the magnetic field by Eq. (26), 
so that the radial force is
\beq
\vec{F}_r(r)=\frac{e}{c}\vec{v}\times\vec{B}=-\frac{e^2 \lambda_L}{m_e c^2}H_c^2 e^{2(r-r_0)/\lambda_L} \hat{r}
\eeq
For the radius $r_0$ to expand by $dr_0$ the work that needs to be performed against this
inward radial force is
\beq
dW=\int d^3r n_s F_r(r) dr_0
\eeq
with $n_s$ the density of superconducting carriers. Eq. (58) yields, using Eq. (11),
\beq
dW=\frac{H_c^2}{4\pi} \pi \ell r_0 d(r_0)  =    \frac{H_c^2}{8 \pi} dV
\eeq
which is the energy supplied by the system when a volume element $dV$ of the system condenses into the superconducting state. 

How does the system supply that energy as electrons condense into the superconducting state? 
Carriers condensing from the normal to the superconducting state at the phase boundary suddenly acquire kinetic energy, the kinetic energy
of the supercurrent $n_sk_e$, with $k_e$ given by Eq. (40), which is the condensation free energy Eq. (6). 
As the boundary moves further out, that kinetic energy is gradually transferred to the electromagnetic field through the second term on the right-hand side of Eq. (47), since
the Faraday electric field points in direction opposite to the supercurrent (counterclockwise), decelerates the electrons
carrying the supercurrent  and ultimately stops them when the phase boundary
has moved out beyond several London penetration depths.
That kinetic  energy Eq. (59), which came from Eq. (6), the condensation free energy, gets transferred to the power source that produces the applied magnetic field $H=H_c$
through the current $I$ given by Eq. (1) flowing counterclockwise in a solenoid surrounding the superconducting sample.
 
This completely accounts for  the energetics of the flux expulsion process. But it leaves the following
 questions about the Meissner effect unanswered:  

\begin{enumerate}
\item{ How does the current acquire its azimuthal momentum when a domain is born?}

\item{How does the condensation energy become  kinetic energy of the supercurrent?}

\item{ How is  azimuthal momentum transferred between the supercurrent and the body as a whole so that angular momentum conservation is
not violated?}

\item{ How does a domain grow outward, overcoming the Maxwell pressure that
wants to  shrink it?}

\item{  How does the azimuthal clockwise current  initially overcome the counterclockwise Faraday electric field
that eventually stops it?}

\item{  How does all of the above happens without energy dissipation to ensure the reversibility of the
process?}

\end {enumerate}
And of course the same questions in reverse exist and need to be answered about the reverse process, the superconductor to normal transition in a magnetic field \cite{scstops}. 
The conventional theory of superconductivity  has not addressed any of these questions. They will be
addressed in the following sections.

 \section{Thermodynamic forces versus real forces, emergence versus reductionism}
 BCS theory has not addressed the questions discussed in the previous sections in the scientific literature. 
 Possible answers to some of  these questions within BCS theory have been gathered by this author from private communications with colleagues. 
 Regarding energy, BCS will say that the theory satisfies energy conservation, and questions such as how does kinetic energy get converted
 into binding energy of Cooper pairs and vice versa do not need to be answered. In the Meissner effect, the system will find its way to its lowest energy state with
 the magnetic field excluded through `fluctuations', regardless of the opposing Faraday field. Regarding momentum, BCS will say that there are
 ``thermodynamic forces'' that will give or take momentum to or from the electrons as they enter or leave the superconducting state
 in order  to drive the system  to its  lower energy states, and that the ions
will acquire compensating angular momentum because if they didn't momentum conservation would be violated.
 And, that the processes may involve collisions, but if those collisions are elastic they would not generate Joule heat and not spoil reversibility.
 And, that in any event when phase boundaries move entropy is necessarily generated so the issue of reversibility is moot \cite{goldenfeld2}.
  And, that superconductivity is an `emergent property' \cite{emergent,anderson}, hence seeking a reductionist answer that tries to identify
 microscopic mechanisms to explain these puzzles is fruitless.

 Whether those answers are true or false is unknown. According to the general consensus \cite{bcs50}, they are true and no further explanation of the issues discussed is
 needed \cite{needed}.  In this author's view, those answers are false \cite{meissnerpapers}. `Thermodynamic forces' ultimately arise from real forces, and it should be possible to
 identify the real forces at play in giving rise to the puzzling momentum transfer phenomena observed. Answers to those questions involving
 real forces have been proposed within the theory of hole superconductivity \cite{holesc,book}. 
  
 It is generally said that there are only four forces in nature,  namely gravitational, electromagnetic, weak and strong. 
 Of those, clearly only the electromagnetic force can play a role. There is however also a
 fifth real  force in nature, namely quantum pressure \cite{emf}. Quantum pressure is
 the fact that a quantum particle will lower its kinetic energy if its wavefunction expands. A particle in a box exerts a force against
 the walls because moving the walls outward will lower the quantum kinetic energy of the particle. Within the theory of hole superconductivity,
 answers to these questions  result from action of the electromagnetic force and this quantum force.  Whether those answers are true or false is unknown. 
 According to the current general consensus they are false, but no specific reasons have been given for why they are false to this author's knowledge.

            \begin{figure} [t]
 \resizebox{8.5cm}{!}{\includegraphics[width=6cm]{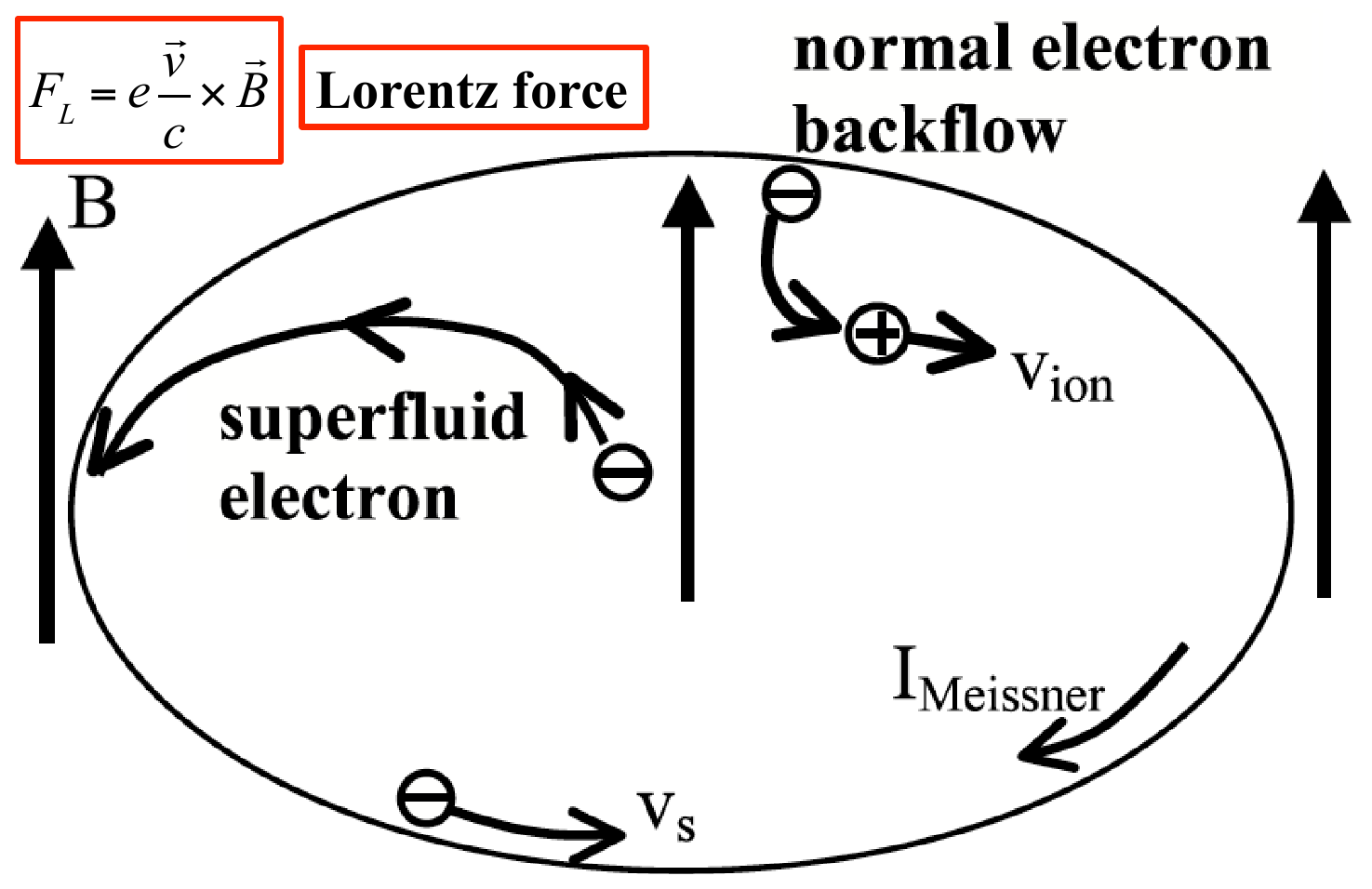}} 
 \caption {Schematic depiction of radial motion of charge to explain the Meissner effect.
 Negative charge moving radially outward acquires counterclockwise azimuthal momentum through
 the Lorentz force, giving rise to the clockwise Meissner current that expels the magnetic field.
 Negative charge backflowing inward get clockwise azimuthal momentum through the Lorentz force
 and transfer that momentum to the ions, by a process explained later, thus ensuring momentum
 conservation. }
 \label{figure1}
 \end{figure} 
 
 \section{Overview of  a proposed reductionist explanation}
 The key fact that can provide an explanation to all the puzzles described, that is predicted to occur   \cite{chargeimbalance} within the
 theory of hole superconductivity, is: {\it radial charge motion has to be involved}, as illustrated in Fig. 7. Qualitatively this is easy to understand.
 Electrons moving radially $outward$, under the influence of the  Lorentz force
 \beq
 \vec{F_L}=\frac{e}{c}\vec{v}\times\vec{B}
 \eeq
 will acquire azimuthal momentum such that the resulting current opposes the applied magnetic field, explaining the expulsion of magnetic field
 that is the Meissner effect \cite{lorentz}. Backflowing electrons flowing radially inward will acquire azimuthal momentum in opposite direction,
 and if they transfer this azimuthal momentum to the body as a whole it would not cancel the magnetic field expulsion but will
 cancel the mechanical momentum of the Meissner current, ensuring momentum conservation \cite{missing}. 
 The driving force for radial outflow of electrons is quantum pressure of electrons going from the normal to the
 superconducting state, originating in quantum kinetic energy lowering, the driving force for backflow of normal electrons that transfer their azimuthal momentum to the lattice is
 the electric force resulting from the electric potential difference created by the radial outflow. 
 The transfer of azimuthal momentum from backflowing electrons to the body occurs reversibly if the backflowing normal electrons
 have negative effective mass, hence are hole carriers \cite{momentum}, as explained in the next sections.
 
            \begin{figure} [t]
 \resizebox{8.5cm}{!}{\includegraphics[width=6cm]{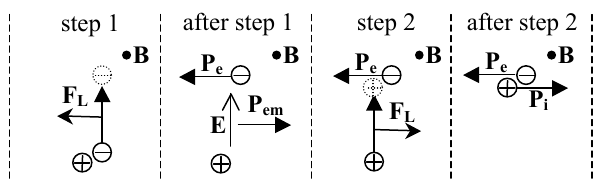}} 
 \caption { Conceptual illustration of how momentum transfer between electrons and ions mediated by the electromagnetic field occurs in the Meissner effect. Magnetic field points out of the picture. We start with a charge-neutral system with one electron and one positive
 charge.  In step 1, electron moving up
 acquires mechanical momentum  \boldmath$P_e$ pointing left through the Lorentz force. Electric field {\bf E} pointing up results from this charge
 motion, which gives rise to electromagnetic momentum $P_{em}$  pointing right ( Eq. (61) ) that exactly cancels \boldmath$P_e$ \unboldmath. In step 2, positive charge moves up the same distance and
 acquires mechanical momentum to the right through the Lorentz force. The electromagnetic momentum
 disappears, and positive and negative charges acquired opposite momentum without any scattering
processes  involved.}
 \label{figure1}
 \end{figure} 
 
The key reason why scattering and dissipation is not involved in these processes is that the transfer of momentum is mediated by the 
electromagnetic field. That is what allows the kinetic energy of the supercurrent to remain  in the electronic degrees of freedom, where it is used to pay the price of the
condensation energy in rendering the superconducting electrons normal, and yet its momentum to be transferred to the ionic
degrees of freedom.

In most physical interactions, momentum transfer is accompanied by energy transfer. An exception is when magnetic fields are
involved. A charge moving in a magnetic field will change its momentum but not its kinetic energy: the magnetic field doesn't do work
on moving charges since the magnetic Lorentz force $\propto \vec{v}\times\vec{B}$   is perpendicular to the particle's velocity $\vec{v}$. The momentum change of the particle is
compensated by momentum change of the electromagnetic field. The momentum density of the electromagnetic field is given by
\beq
\vec{\mathcal{P}}_{em}(\vec{r})=\frac{1}{4\pi c} \vec{E}\times\vec{B} 
\eeq
with $\vec{E}, \vec{B}$ electric and magnetic fields. In order for the electromagnetic momentum Eq. (61) to be azimuthal, and thus be able to 
mediate the azimuthal momentum transfer between electrons and ions, requires that the electric field $\vec{E}$ in Eq. (61) is $radial$.
To generate radial electric fields requires radial charge flow. 

Let us understand conceptually how this transfer of momentum between electrons and ions mediated by the electromagnetic field can happen in the Meissner effect, with reference to Fig. 8. The up direction corresponds
to radial outward motion of the superconductor-normal phase boundary. If the outward motion
of the phase boundary is associated with outward motion of negative charge, the negative charge
acquires momentum to the left (counterclockwise) through the Lorentz force, and this  creates a transitory radially $outgoing$ (upward in Fig. 8)  electric field
and hence a clockwise electromagnetic field momentum that  compensates the counterclockwise mechanical momentum acquired by the
outward-moving electron due to the Lorentz force. In the subsequent step 2,  positive charge moves 
up (radially outward) and 
the clockwise momentum of the electromagnetic field is transferred to the positive charge through the Lorentz force. The end result is negative and
positive charges moving with the same momentum in opposite directions, as shown   in the rightmost panel of Fig. 8. 
Similarly the processes that transfer the momentum of the supercurrent to positive charges in the superconductor to normal transition are shown in Fig. 9.

If the positive charges in Figs. 8 and 9 are ions, the processes explains how ions  acquire
the compensating momenta. However, of course ions in a solid cannot move radially outward nor inward.
Instead, the positive charges in Figs. 8 and 9 are $holes$, or equivalently  electrons with negative
effective mass. The process of flux expulsion involves electrons becoming superconducting
moving outward and at the same time normal holes moving outward to compensate for
the charge imbalance and transfer momentum to the ions without scattering processes  \cite{momentum}.
 For the superconductor to normal transition the
same processes occur, in reverse. 

            \begin{figure} [t]
 \resizebox{8.5cm}{!}{\includegraphics[width=6cm]{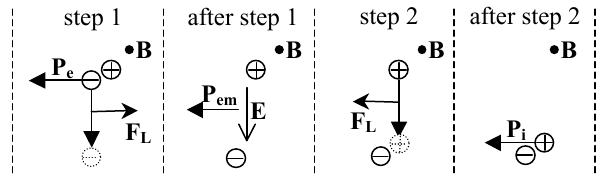}} 
 \caption { Same as Fig. 8 illustrating how the supercurrent stops and its momentum is transferred to positive charges
 in the superconductor to normal transition.}
 \label{figure1}
 \end{figure} 

\section{Charge expulsion and Alfven's theorem}
The fact that the Meissner effect, which involves outward motion of magnetic field lines, would be associated with
outward motion of charge should not be surprising. That is precisely what classical plasmas
would do. While the ultimate laws governing superconductivity operate in the quantum realm,
superconductivity is a macroscopic phenomenon and Bohr's correspondence principle teaches us
that there should be a smooth connection between classical and quantum realms.
It is well known in the study of magnetohydrodynamics of conducting fluids that motion of magnetic 
field lines is closely associated with motion of charge \cite{davidson}. This is codified precisely in 
Alfven's theorem \cite{alfventhm}: in a perfectly conducting fluid, magnetic field lines are frozen with the fluid and
move together with the fluid. Of course this is a necessary consequence of Faraday's law.
If the fluid is less than perfectly conducting there will be some relative motion of fluid and magnetic field 
lines, still, as P. H. Roberts points out \cite{roberts}, {\it ``Alfven's theorem is also helpful in
attacking the problem of inferring unobservable  fluid motions from observed magnetic field behavior''}. For example, measurements of magnetic  field variations near
one of Jupiter's moons demonstrated the existence of an unobservable conducting fluid below its surface
\cite{kivelsonmg}.
Isn't  it reasonable to expect that the observed motion of magnetic field lines in the Meissner effect should allow us to infer {\it `unobservable
fluid  motions'} underlying it? \cite{alfvenmine}

Fig. 10 shows schematically the behavior of magnetic field lines resulting from jets of
conducting fluid flowing from the center. They resemble the behavior of magnetic field lines
in the Meissner effect, Fig. 1. In a classical plasma such fluid  flow would not occur, since it would
give rise to a void at the center. It requires outward flow of a fluid that carries neither net
charge nor net mass. In a metal going superconducting it can occur if what flows out
is negative electrons and positive holes, since holes flowing out is associated with 
real mass (electron mass) flowing in.

           \begin{figure} [t]
 \resizebox{8.5cm}{!}{\includegraphics[width=6cm]{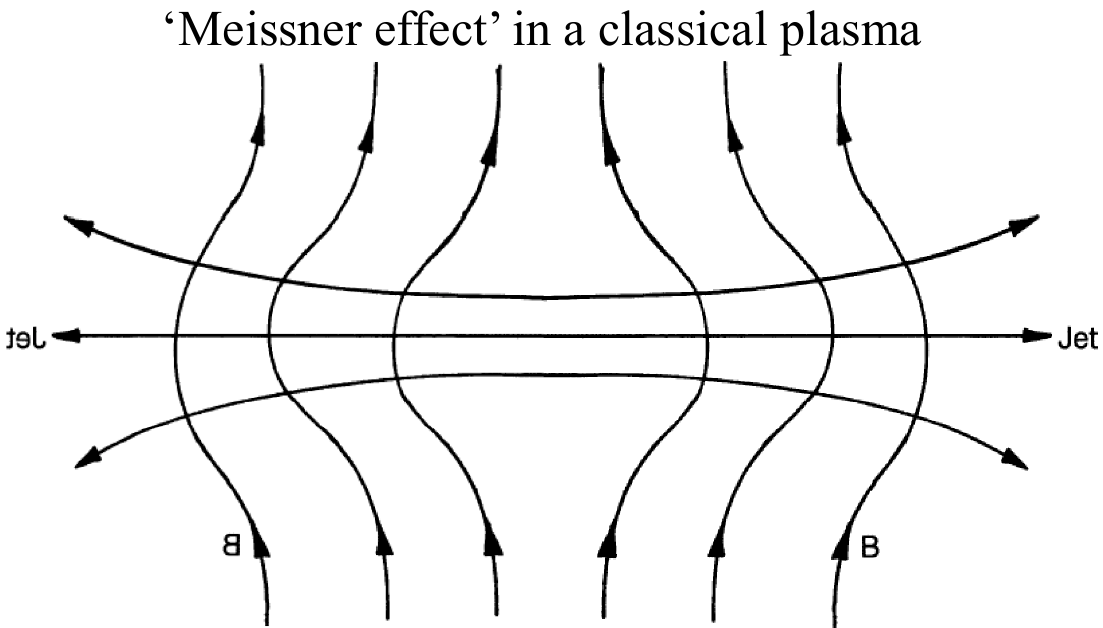}} 
 \caption {The right half of this figure is Fig. 2.8 of Ref. \cite{davidson}, ``An Introduction to
 Magnetohydrodynamics'', with caption 
 {\it ``An example of Alfven's theorem: flow through a magnetic field causes the field lines to bow out''},
 illustrating the bending of magnetic field lines by a jet of conducting fluid. The left half of the figure
 was copied  from the right side and flipped horizontally. }
  \label{figure1}
 \end{figure} 

 Let us consider again the treatment of London electrodynamics of Sect. VI, where some terms were neglected. 
 Rather than Eq. (7), the more accurate equation is
\beq
\frac{d\vec{v}}{dt}=\frac{e}{m_e}\vec{E}+\frac{e}{m_ec}\vec{v}\times\vec{B}
\eeq
The left-hand side of Eq. (62) is the total (convective) time derivative, which is related to the local (partial) time derivative by 
\beq
\frac{d\vec{v}}{dt}=\frac{\partial \vec{v}}{\partial t} + (\vec{v}\cdot \vec{\nabla})\vec{v}= \frac{\partial \vec{v}}{\partial t} +\vec{\nabla}(\frac{\vec{v}^2}{2})-\vec{v}\times(\vec{\nabla}\times\vec{v})
\eeq
Defining the  `generalized vorticity'
\beq
\vec{w}=\vec{\nabla}\times\vec{v}+\frac{e}{m_ec}\vec{B},
\eeq
taking the curl of Eq. (63) and using Eq. (64) and Faraday's law $\vec{\nabla} \times \vec{E}=-(1/c)\partial \vec{B}/\partial t$  leads to the following equation of motion for $\vec{w}$ \cite{londonbook}:
\beq
\frac{\partial \vec{w}}{\partial t} = \vec{\nabla}\times(\vec{v}\times\vec{w})  .
\eeq

Note that $\vec{w}$ is essentially the curl of the canonical momentum $\vec{p}=m_e \vec{v}+(e/c)\vec{A}$, with $\vec{A}$ the magnetic vector potential. 
In the Meissner process we have  at time $t=0$:
$
\vec{w}(\vec{r},t=0)= e/(m_ec)\vec{B}(t=0)\equiv \vec{w}_0 \neq 0
$
independent of position $\vec{r}$. We set $\vec{\nabla}\times\vec{v}=0$ because in the normal state there is no net macroscopic charge flow.
Hence the canonical momentum $\vec{p}$ is nonzero throughout the interior of the superconductor in the initial state. 
In the superconducting state, the superfluid velocity $\vec{v}$ obeys the London equation 
$
\vec{\nabla}\times\vec{v}=-e/(m_ec)\vec{B}. 
$
Therefore, $\vec{w}(\vec{r},t=\infty)=0$
everywhere in the superconducting body. 
Equivalently,  the canonical momentum $\vec{p}=0$ throughout the interior of the (simply connected)  superconductor.
In a cylindrical geometry, assuming azimuthal symmetry as well as translational symmetry along the cylinder axis ($z$) direction (infinitely long cylinder)
$\vec{w}(\vec{r},t)=w(r,t)\hat{z}$
and Eq. (65) takes the form
\beq
\frac{\partial w}{\partial t} =-\frac{1}{r}\frac{\partial}{\partial r}(rwv_r)
\eeq
with $r$ the radius in cylindrical coordinates. Eq. (66)  implies that  {\it  $ w $ can only change if there is radial flow of charge} ($v_r \neq 0$). Moreover,  for $w$ to evolve towards
its final value $0$ requires $v_r>0$, i.e. a radial $outflow$ of electrons.

\section{ Momentum transfers by radial motion}
How much charge needs to flow radially out  and how far to explain the Meissner effect? The answer is simple \cite{reversannals}.  The velocity of electrons in the Meissner current for magnetic field $B=H_c$ at the phase boundary is
\beq
v_s=-\frac{e\lambda_L}{m_e c}H_c .
\eeq
While Eq. (27) gave the more detailed radial dependence of the Meissner velocity, it is equivalent to saying that all electrons within distance $\lambda_L$ of the surface or of the phase boundary between superconducting and normal phases are moving with
velocity Eq. (67).
If in the process of joining the condensate normal electrons form Cooper pairs $and$ move radially outward, as shown schematically in
Fig. 11, they will acquire through
the Lorentz force  Eq. (60)  azimuthal velocity $v_\theta$ given by
\beq
\frac{d v_\theta}{dt}=\frac{e}{m_e c}\frac{dr}{dt}B
\eeq
hence
\beq
\Delta  v_\theta=\frac{e}{m_e c} \Delta r B
\eeq
Therefore, if initially the electron had zero   azimuthal velocity on average, by moving radially outward a distance $\Delta r = \lambda_L$ it will acquire the azimuthal momentum  required by the Meissner
velocity Eq. (67):
\beq
\Delta p=\frac{e \lambda_L}{c} H_c .
\eeq
The outward motion needs to be sufficiently fast so that we can ignore the azimuthal momentum in the opposite direction
imparted by the Faraday electric field during this process.

           \begin{figure} [t]
 \resizebox{8.5cm}{!}{\includegraphics[width=6cm]{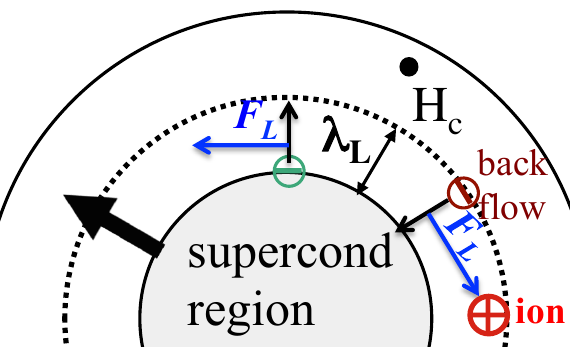}} 
 \caption { Schematic depiction of momentum transfers in the Meissner effect. Green electron moving radially outward  a distance $\lambda_L$ acquires through the Lorentz force counterclockwise azimuthal momentum with velocity equal to the  azimuthal velocity of
 electrons in the Meissner current, Eq. (53).  Backflowing brown electron   acquires the same azimuthal momentum in clockwise direction through the
 Lorentz force and
 transfers it to the ions. }
  \label{figure1}
 \end{figure} 

To compensate for this change in azimuthal momentum, as well as to preserve charge homogeneity, there needs to be a backflow of normal electrons
that acquire the same azimuthal momentum in clockwise direction through the Lorentz force,  $and$ transfer that momentum to the body
as a whole, as also shown schematically in Fig. 11. 

In summary, the processes shown in Fig. 11 explain qualitatively how the puzzling momentum transfers happen. Each electron when
transiting from the normal to the superconducting state sprints radially outward a distance $\lambda_L$ and in the process acquires 
the azimuthal momentum needed to contribute its share to the Meissner current. And for each such outflow there is a backflowing electron that acquires the same azimuthal momentum in the opposite
direction and transfers it to the body.

How this happens in detail is discussed in the following sections.

           \begin{figure} [t]
 \resizebox{8.5cm}{!}{\includegraphics[width=6cm]{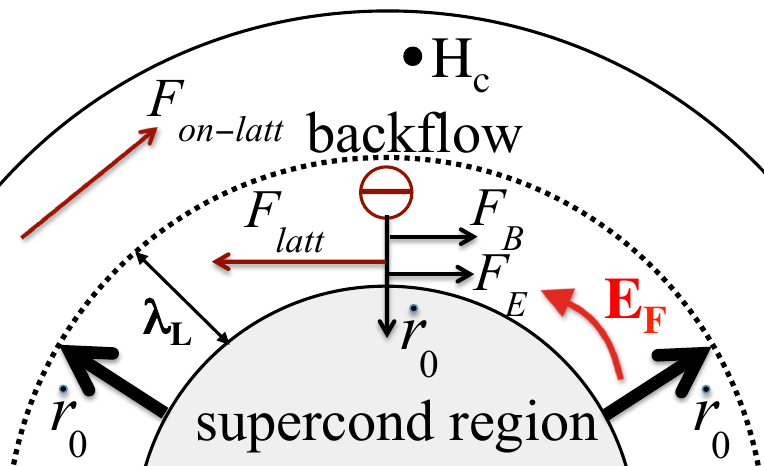}} 
 \caption {How normal electrons backflowing into the phase boundary with speed $\dot{r}_0$ transfer azimuthal momentum to the lattice
 without scattering processes in the Meissner effect.
 $F_E$ is the electric force due to the Faraday field $E_F$, $F_B$ is the magnetic Lorentz force for speed of motion $\dot{r}_0$,
 and $F_{latt}=F_E+F_B=2F_E=2F_B$ is the force exerted by the lattice on the electron. $F_{on-latt}=F_{latt}$ is the reaction force of
 the electron on the body.  }
  \label{figure1}
 \end{figure} 
 
 \section{ Momentum transfer to the ions}
 We first discuss how backflowing normal electrons transfer azimuthal momentum to the body without collisions.
 The relevant forces are shown in Fig. 12. The Faraday electric force is 
 \beq
 F_E=eE_F=e\frac{\dot{r}_0}{c}H_c
 \eeq
 and it is identical to the magnetic Lorentz force
  \beq
 F_B=e\frac{\dot{r}_0}{c}H_c 
 \eeq
 imparted to normal electrons backflowing with speed $\dot{r}_0$. Both forces
  point clockwise. Just like for electrons in a Hall bar with positive Hall coefficient, the electromagnetic forces are balanced
 by the force exerted by the lattice on the electron, $F_{latt}$, provided the electron has negative effective mass \cite{disapp,momentum}:
 \beq
 F_{latt}=2e\frac{\dot{r}_0}{c}H_c = F_{on-latt} .
 \eeq
 By Newton's third law, the electrons exert equal and opposite force on the lattice, i.e. in clockwise direction, $F_{on-latt}$.
 The momentum that is transferred to the lattice by the normal electron backflowing a radial distance $\lambda_L$ in time
 $\Delta t$, with $\dot{r}_0=\lambda_L/\Delta t$  is then
 \beq
 \Delta p_{on-latt}=F_{on-latt} \Delta t=2e\frac{\dot{r}_0}{c}H_c  \Delta t = 2\frac{e\lambda_L}{c}H_c .
 \eeq
 In addition the Faraday electric field, by exerting force on the positive ions transfers half of that momentum in counterclockwise 
 direction, hence the net transfer of momentum to the body is
  \beq
 \Delta p_{on-latt}^{net}= \frac{e\lambda_L}{c}H_c .
 \eeq
 identical to the momentum in counterclockwise direction acquired by the outflowing electrons, Eq. (70).
 
 This requires that backflowing electrons have negative effective mass, as explained in   detail in the references  \cite{disapp,momentum,reversannals}.

            \begin{figure} [t]
 \resizebox{8.5cm}{!}{\includegraphics[width=6cm]{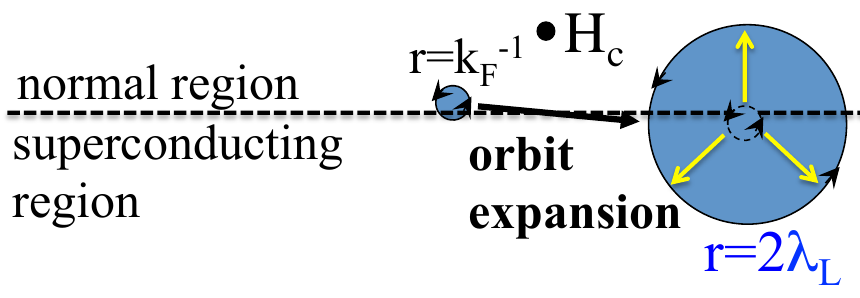}} 
 \caption { Normal electrons with orbits of microscopic radius expand their orbits to radius $2\lambda_L$ as they
 enter the superconducting region. Through the action of the Lorentz force on the expanding orbit they
 acquire the azimuthal velocity of the Meissner current Eq. (67) }
  \label{figure1}
 \end{figure}

\section{ Momentum transfer to the electrons}
Assume that in the process of forming Cooper pairs and becoming superconducting, electrons expand their orbits from
a microscopic radius of order  $k_F^{-1}$, with $k_F$ the Fermi wavevector, to mesoscopic radius $2\lambda_L$ \cite{sm}.
We will discuss in the next section why this is reasonable. If that happens, when electrons at the phase boundary expand their orbits
they will expand into the normal region, as shown in Fig. 13. Through the action of the Lorentz force on the expanding
orbit, the azimuthal velocity that the electron acquires is precisely the speed of the Meissner current, Eq. (67).
This  is seen as follows: the equation of motion as the electron moves radial outward  is
\beq
m_e\frac{d\vec{v}}{dt}=\frac{e}{c}\vec{v}\times\vec{B}+\vec{F}_r
\eeq
where the first term is the magnetic Lorentz force and we have included a radial force arising from ``quantum pressure'' that drives the orbit expansion.
From Eq. (76) we infer
\beq
\vec{r}\times\frac{d\vec{v}}{dt}=\frac{e}{m_ec}\vec{r}\times (\vec{v}\times\vec{B})
\eeq
where $\vec{r}$ is in the plane perpendicular to the axis of the cylinder. Hence $\vec{r}\cdot\vec{B}=0$ and 
$\vec{r}\times (\vec{v}\times\vec{B})=-(\vec{r}\cdot\vec{v})\vec{B}$, and
\beq
\frac{d}{dt}(\vec{r}\times\vec{v})=-\frac{e}{m_ec}(\vec{r}\cdot\vec{v})\vec{B}=-\frac{e}{2m_ec}(\frac{d}{dt}r^2)\vec{B}
\eeq
so that $\vec{r}\times\vec{v}=-(e/2m_e c)r^2\vec{B}$, and the acquired azimuthal velocity in moving out a distance $r$ is
\beq
v_\phi=-\frac{e}{2m_ec}rB
\eeq
Thus,  the electron acquires azimuthal speed $v_s=-e\lambda_L/(m_e c)B$ in expanding its orbit to  radius $r=2\lambda_L$. 
In the region right outside the phase boundary the expanded orbits overlap the microscopic orbits of normal electrons,
as seen in Fig. 14, and the extra negative charge induces the inward  `backflow' of normal electrons.

            \begin{figure} [t]
 \resizebox{6.5cm}{!}{\includegraphics[width=6cm]{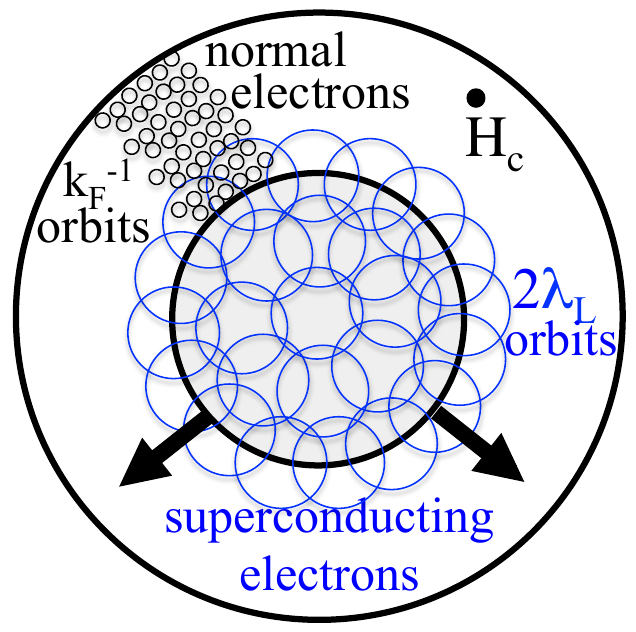}} 
 \caption { Superconducting electrons reside in overlapping mesoscopic orbits of radius $2\lambda_L$. In the normal
 region, electrons reside in non-overlapping microscopic orbits of radius $k_F^{-1}$. 
 The negative charge from the expanded orbits extends into the normal region and induces the `backflow' of
 normal electrons shown in Figs. 11 and 12. }
  \label{figure1}
 \end{figure} 
 
 The above provides a dynamical explanation for how electrons acquire the azimuthal velocity of the Meissner current
 as they enter the superconducting state. In the next section we present several arguments in favor of this hypothesis.

\section{ Mesoscopic orbits}
 The
Larmor diamagnetic susceptibility for carriers of density $n$ in orbits of radius r is \cite{am}
\beq
\chi_{Larmor}(r)=-\frac{ne^2}{4m_e c^2}r^2
\eeq
It is remarkable that the normal state Landau diamagnetism of metals  is described by this expression for normal state orbits of microscopic radius $k_F^{-1}$:
\beq
\chi_{Landau}=-\frac{1}{3} \mu_B^2 g( \epsilon _F)=-\frac{ne^2}{4m_e c^2}k_F^{-2}=\chi_{Larmor}(k_F^{-1})
\eeq
with $k_F$ the Fermi wavevector, $g(\epsilon_F)=3n/2\epsilon_F$ the free electron density of states, and $\mu_B=e\hbar/2m_ec$ the Bohr magneton.
It is also remarkable that the perfect diamagnetism of the superconducting state is described by the Larmor formula Eq. (80) for orbits of mesoscopic radius $2\lambda_L$:
\beq
\chi_{London}=-\frac{1}{4\pi}=-\frac{ne^2}{4m_e c^2}(2\lambda_L)^2=\chi_{Larmor}(2\lambda_L)
\eeq
with  the London penetration depth $\lambda_L$ given by the usual form Eq. (11) 
and $n_s=n$ the superfluid density. This indicates that (i) the transition to superconductivity can be understood as involving
an $expansion$ of electronic orbits from radius $k_F^{-1}$ to radius $2\lambda_L$, and (ii) that the superconducting ground state
can be understood as being composed of  all electrons ($n_s=n$) residing in orbits of radius $2\lambda_L$ \cite{sm}.

This is further 
supported by the observation that the  angular momentum carried by the electrons in the Meissner current 
with velocity $v_s$ in a cylinder of radius $R>>\lambda_L$
and height $\ell$ can be written in the  two equivalent forms:
\beqn
L_{Meissner}&=&n_s (2\pi R \lambda_L \ell)\times (m_e v_s R) \nonumber \\
&=& n_s (\pi R^2 \ell)\times(m_e v_s (2\lambda_L))
\eeqn
The first form is the conventional description of electrons flowing in a surface layer of thickness $\lambda_L$, hence cross-sectional area $2\pi R\lambda_L$, each moving in 
a circle of radius $R$. 
The second form describes all electrons in the bulk, hence cross-sectional area $\pi R^2$, each moving in a mesoscopic orbit of radius $2\lambda_L$. This is shown schematically in Fig. 15.

            \begin{figure} [t]
 \resizebox{8.5cm}{!}{\includegraphics[width=6cm]{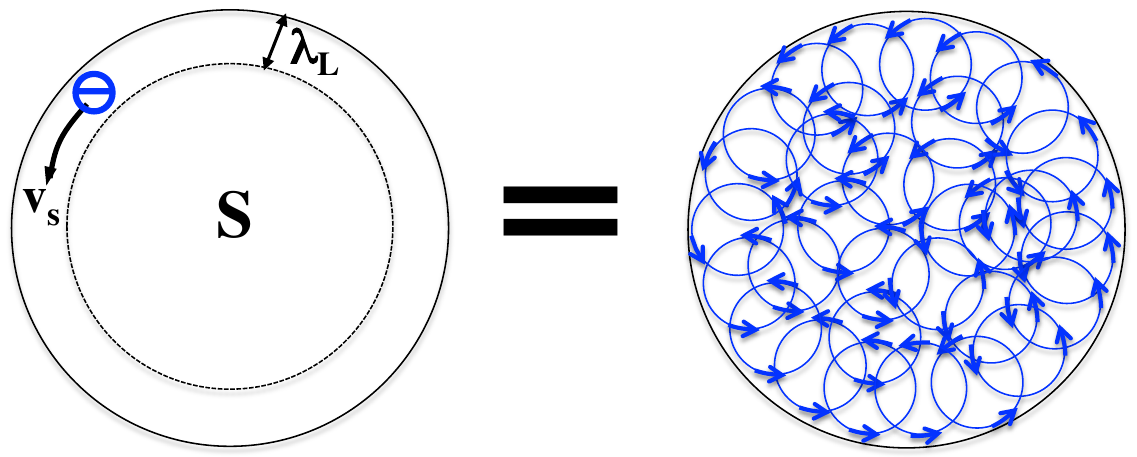}} 
 \caption { The Meissner current resulting from carriers within a shell of thickness $\lambda_L$ from the surface
 circulating with speed $v_s$ (left panel) can be regarded as arising from individual electrons orbiting with the
 same speed in overlapping orbits of radius $2\lambda_L$, as shown by Eq. (62).}
  \label{figure1}
 \end{figure}

           \begin{figure} [t]
 \resizebox{5.5cm}{!}{\includegraphics[width=6cm]{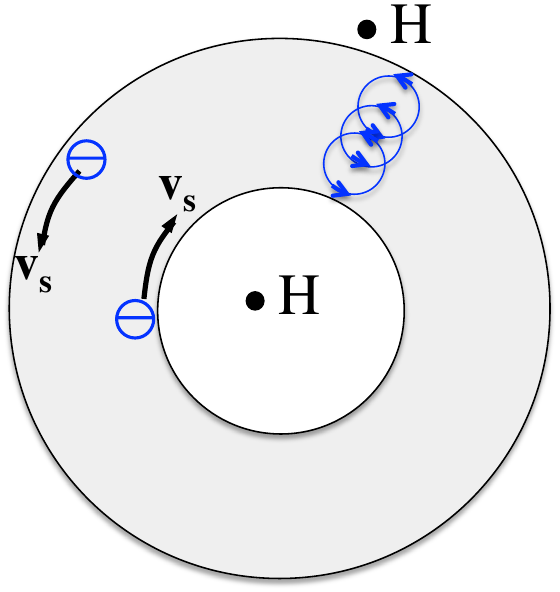}} 
 \caption {Cylindrical shell cooled into the superconducting state in the presence of a magnetic field as seen from the top.  Currents flow
 along the outer and inner surfaces in opposite directions. This is a natural consequence of 
 the orbits  expanding in a magnetic field and acquiring counterclockwise velocity through the Lorentz force.}
  \label{figure1}
 \end{figure}

This concept is also helpful for understanding what happens when a cylindrical shell is cooled in a magnetic field, as shown in Fig. 16.
The magnetic flux in the empty region inside the shell remains essentially unchanged,
up to at most a fraction of a flux quantum which we ignore. 
During the process, a clockwise Faraday electric field exists opposing the expulsion of flux from the shell.
The field is expelled from the shell region, which requires that supercurrents flow in opposite directions along the
outer and inner surfaces of the shell. Within the `emergence' viewpoint, there is no explanation for how electrons acquire their
momentum in opposite directions in the outer and inner surfaces, it   follows from the equations
and the fact that the system ends up in the lowest energy state.
Instead, within our `reductionist' point of view, the fact that electrons acquire opposite velocities in the outer and
inner surfaces  is a natural consequence of the expansion of microscopic orbits to radius $2\lambda_L$, as shown in Fig. 16.

            \begin{figure} [t]
 \resizebox{8.5cm}{!}{\includegraphics[width=6cm]{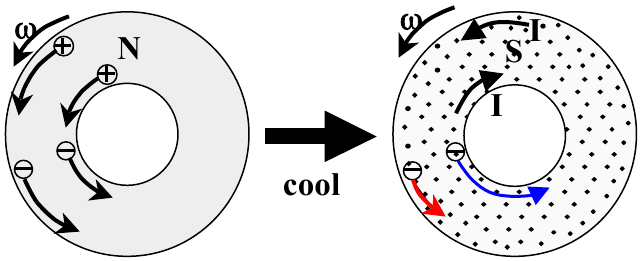}} 
 \caption {Rotating cylindrical shell that is cooled into the superconducting state while rotating.
 In the normal state, electrons and ions rotate with the same tangential speed $\omega r$.
 A uniform magnetic field develops in the interior. This means that electrons
 near the outer surface need to slow down and electrons near the inner surface need to speed up
 relative to their motions in the normal state.  }
  \label{figure1}
 \end{figure}

\section{ Rotating superconductors}

Another situation related to the Meissner effect occurs when a normal metal that is rotating is cooled
into the superconducting state, in the absence of applied field. This is called the London moment. 
Fig. 17 shows what happens for a rotating cylindrical shell \cite{rotatingexperiment}. 
A uniform magnetic field develops in the interior of the shell, which implies that currents 
are generated that circulate
in opposite directions near the outer and inner shell surfaces. 
When the system is rotating in the normal state, electrons and ions  move at the same tangential speed given
by $\omega r$, with $\omega $ the angular velocity and $r$ the radial position of the ion or electron.
For the magnetic field to develop in the interior of the shell, in the transition to superconductivity the
electrons near the outer surface need to slow down and those the inner surface need to speed up,
as shown in Fig. 17

This behavior is easy to understand with $2\lambda_L$ orbits \cite{rotating}, as shown in Fig. 18.
In the normal state we can think of the electron at radius $r$ as a point particle, with moment of inertia
$m_e r^2$ and angular momentum $l_e=m_e r^2 \omega$. When its orbit expands to radius $2\lambda_L$ its angular momentum cannot change.
That means that it has to start rotating in clockwise direction with angular velocity 
$\omega$ in the rest frame of the rotating cylinder, as indicated by the red arrow in Fig. 18, since
\beq
l_e=m_e r^2 \omega=m_e(r^2+(2\lambda_L)^2)\omega -m_e(2\lambda_L)^2\omega .
\eeq
Superposition of the resulting rotating orbits, just as shown in Fig. 16 right panel,
 gives rise to slowdown near the outer surface and
speedup near the inner surface, as required by the velocity patterns shown in Fig. 17 right panel.

    \begin{figure} [t]
 \resizebox{8.5cm}{!}{\includegraphics[width=6cm]{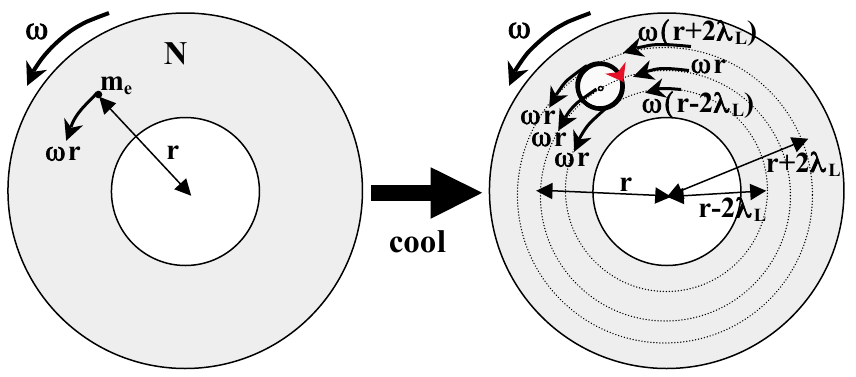}} 
 \caption {Explanation of why electrons speed up and slow down near the inner and upper
surfaces of the shell of Fig. 17. Electrons enlarge their orbits from a microscopic radius to
 radius $2\lambda_L$, with their total angular momentum unchanged. This implies that they
 rotate clockwise with angular velocity $\omega$ in the rotating frame, as shown by the 
 red arrow on the right panel. }
  \label{figure1}
 \end{figure}

Quantitatively this is understood as follows. The London equation
\beq
\vec{\nabla}\times \vec{v_s}=-\frac{e}{m_ec}\vec{B}
\eeq is assumed to be valid, with $\vec{v}_s$ the velocity of electrons in the non-rotating frame.
Deep in the interior of the shell electrons rotate at the same speed as ions, with velocity
$\vec{v}_s=\vec{\omega}\times\vec{r}$ at radius $\vec{r}$. Since $\vec{\nabla}\times( \omega\times\vec{r}) =2\vec{\omega}$,
 it follows that the magnetic field is
\beq
\vec{B}_0=-\frac{2m_e c}{e}\vec{\omega} .
\eeq
Now at the outer surface, the relative velocity between electrons and the body gives rise to the interior field, hence
\beq
\vec{\nabla}\times(\vec{v}_s-\vec{v}_0)=-\frac{e}{m_e c}\vec{B}_0
\eeq
from which it follows in the usual way
\beq
v_s-v_0=-\frac{e\lambda_L}{m_e c} B_0
\eeq
hence
\beq
v_s-v_0=-2\lambda_L\omega .
\eeq
Eq. (89) says that the slowdown of electrons at the outer surface is identical to what would result if electrons expanded their orbits
to radius $2\lambda_L$ and in the process acquired clockwise angular velocity $\omega$, as shown in Fig. 18.
The same physics leads to, for electrons near the inner surface of the shell
\beq
v_s-v_0=+2\lambda_L\omega .
\eeq
so electrons speed up relative to the ions to cancel the magnetic field in the empty region inside the shell.

Within an ``emergent'' viewpoint, these results are a consequence the equations, and the question of what is the dynamics leading
electrons to slow down and speed up near the surfaces is not asked. Instead, within a ``reductionist'' viewpoint one would
like to understand the dynamic processes involved. Expansion of electronic orbits to radius $2\lambda_L$ provide
an answer that is understandable and quantitatively correct, as discussed above.

  \begin{figure} [t]
 \resizebox{8.5cm}{!}{\includegraphics[width=6cm]{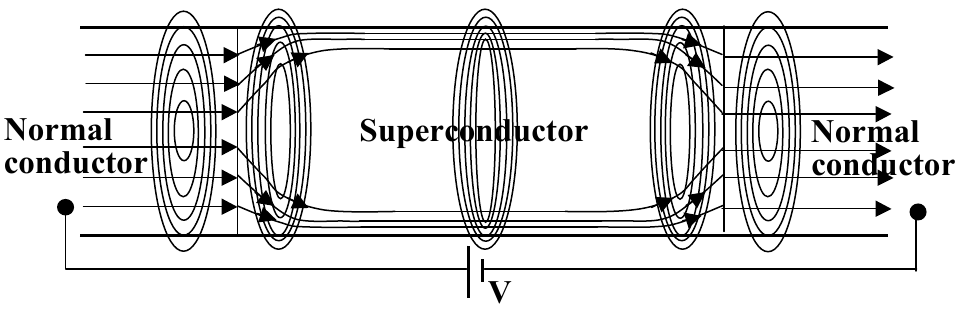}} 
 \caption {Streamlines for electrons entering and leaving the superconducting region of a cylindrical wire.
 The circles indicate magnetic field lines. }
  \label{figure1}
 \end{figure} 
 
\section{ Superconducting wire}
A superconducting wire connected to normal metal leads presents another interesting situation to contrast
emergent and reductionist understandings of the situation.  Fig. 19 show the streamlines and magnetic
field lines for this  case. They are simply obtained from solving London's equation Eq. (10) together with
Ampere's law and  the boundary conditions appropriate to this
situation \cite{londonbook}. The streamlines show a discontinuous slope at the boundaries between normal and
superconducting regions. Within a `reductionist' approach one may wonder, what are the forces that make carriers discontinuously
change their direction when they cross the phase boundary? Instead, within an `emergence' approach, there would be no reason to ask that question.
That is what the London equation predicts, which in turn is predicted by BCS theory, that dictates that
the system knows how to find its way to its lowest energy state, even in a non-static steady state situation.

Consider the cylindrical wire along the z direction, with $-b<z<b$ denoting the superconducting region, and current density
 $J$ flowing in the $-z$ direction, as shown in in Fig. 19 \cite{wire}.
 At the N-S boundary $z=-b$ , the current streamlines acquire {\it discontinuously} motion in the $outward$ radial direction, as seen in Fig. 19. 
 The same is true at the S-N boundary $z=+b$. 
 From the equations and boundary conditions we find that the radial velocity $v_r(r,z)$  acquired at the phase boundary is,
 in terms of the azimuthal magnetic field $B_{\theta}(r,z)$
 \beq
 v_r(r,-b)=-\frac{e\lambda_L}{m_ec}B_{\theta}(r,-b)=\frac{r}{2\lambda_L} v_n
 \eeq
 where $v_n=J/(ne)$ is the carrier velocity in the normal region in direction parallel to the wire, where  $n$ is the density of carriers.
The magnetic field is given by
  \beq
B_{\theta}(r,-b)=\frac{2\pi }{c}Jr  
 \eeq linearly proportional to $r$, since the current density is uniformly distributed across the wire in the normal region.
%
%
Eq. (91)   implies that as carriers enter the superconducting region they suddenly acquire a very large   radial impulse (for $r>>\lambda_L$)
in direction parallel to the phase boundary. Presumably this occurs over a very short time scale,
which implies that an enormous force in the radial direction is acting on the charge carriers as they enter the superconducting region.
                \begin{figure} [t]
 \resizebox{8.5cm}{!}{\includegraphics[width=6cm]{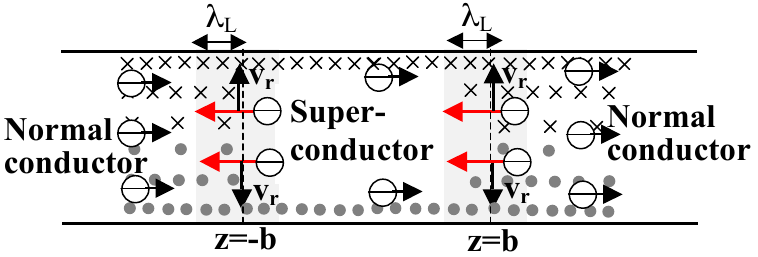}} 
 \caption { In order for carriers to acquire the speed Eq. (74)
in the outward radial direction as they enter or leave the superconducting
region, they have to undergo a sudden motion in the  -z direction
a distance $\lambda_L$ (red arrows). The magnetic field provides the radial impulse through the Lorentz force, it  points into (out of) the paper in the
top (bottom) half of the figure, as indicated by the crosses and circles.  }
  \label{figure1}
 \end{figure} 
Eq. (91) indicates that it is the magnetic field $B_{\theta}$ that imparts the impulse to the carriers in the radial direction as they
become superconducting: the impulse is zero if $B_{\theta}=0$ and it is directly proportional to the local value of $B_{\theta}$ 
for a given $r$, with the same proportionality constant independent of $r$. And the impulse points in direction perpendicular to $\vec{B}$.
It is natural to conclude that the impulse results from the magnetic Lorentz force acting on carriers entering the 
superconducting region:
\beq
\vec{F}_L=\frac{e}{c}\vec{v}\times \vec{B} .
\eeq
 In order for the electron to get an impulse in the radial direction given that $\vec{B}$ is in the azimuthal direction,  $\vec{v}$ in Eq. (93)  has to point in the $-\hat{z}$ direction, as shown in Fig. 20.
 
 The carriers flowing from the normal into the superconducting region acquire the velocity $v_r$ instantly as they cross the phase boundary.
 Let us assume that the instant they cross the phase boundary $z=-b$ they recoil backward (in the $-z$ direction) a distance $\Delta \ell$ in a very short time interval $\Delta t$, so in Eq. (93)
 $v=\Delta \ell/\Delta t$. In order to acquire the speed in the $r$ direction given by Eq. (91) under the action of the Lorentz force
 Eq. (93)  it is necessary that $\Delta \ell=\lambda_L$, so that $F_L\Delta t=(e/c)\Delta \ell B=m_ev_r$.
 Similarly,  when charges leave the condensate at $z=+b$, they also have to acquire a sudden impulse in the $outward$ radial
  direction   of the same magnitude as when they entered  (see streamlines in Fig. 19 near $z=b$). 
 Again this would result  if they move backward (in the $-z$ direction) a distance  $\Delta \ell=\lambda_L$.

 We can understand this behavior with the same concept of mesoscopic orbits discussed earlier, illustrated in Fig. 21. When normal carriers enter
 the superconducting region   their
 orbits expand from  microscopic radius to radius $2\lambda_L$. Just like we discussed earlier in connection with  the momentum transfers in 
 the Meissner effect, the same momentum is acquired through the Lorentz force if the motion is
 linear over a distance $\lambda_L$ or if the orbit expands to radius $2\lambda_L$. 
 In the presence
of a magnetic field, carriers acquire angular velocity that
gives rise to the tangential velocity given by Eq. (91).
Conversely, when pairs leave the condensate their orbits
contract and their tangential velocity goes to zero, hence the motion for $z>b$ is parallel to  the $z$ direction.

             \begin{figure} []
 \resizebox{8.5cm}{!}{\includegraphics[width=6cm]{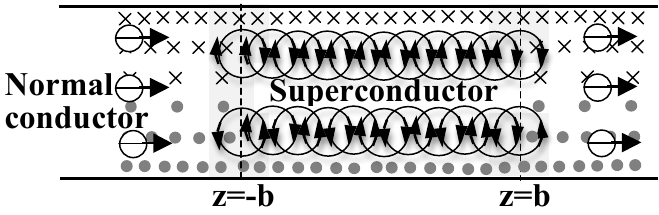}} 
 \caption { The processes of orbit expansion and contraction as carriers enter and leave the superconducting region
explain the sudden impulses in the $-z$ direction and resulting radial impulses acquired by the carriers at the boundaries shown in Fig. 20. }
  \label{figure1}
 \end{figure} 
%
%
%
%
%
%

\section{Theory of hole superconductivity}

The `reductionist' explanations of momentum transfers associated with the  Meissner effect and related phenomena discussed in the earlier sections were developed within the alternative theory of hole superconductivity \cite{holesc,book}. We summarize here the key elements of the theory.

(1) The theory predicted from the outset \cite{hole1} that superconductivity requires hole carriers.  

(2) There is a fundamental asymmetry between carriers near the bottom and near the top of electronic energy bands in solids \cite{dynhub,ehasym}.
This gives rise to a term in the Coulomb interaction between electrons that breaks electron-hole symmetry that becomes increasingly
attractive as the Fermi level approaches the top of the band, that gives rise to pairing and superconductivity \cite{hm89,bondc}.

(3) The attractive interaction lowers the kinetic energy of pairs \cite{hmlondon}, so  superconductivity is driven by lowering of 
kinetic energy rather than potential energy as in BCS.

(4) Kinetic energy lowering drives the orbit expansion discussed earlier \cite{kinenergy}. The quantum kinetic energy
of an electronic wavefunction of radius $r$ is $\hbar ^2/(m_er^2)$ and it decreases as $r$ increases.

(5) The effective mass of normal carriers at the Fermi energy is negative because the Fermi level is close to the
top of the band. As discussed earlier this is essential to explain the transfer of momentum to ions without dissipation.

(6) In the condensate, carriers occupy overlapping mesoscopic orbits of radius $2\lambda_L$, and 
the two members of a Cooper pair orbit in opposite direction, with zero-point velocity \cite{sm}
\beq
v_{\sigma}^0=\frac{\hbar}{4m_e \lambda_L} .
\eeq
Hence their orbital angular momentum is $\hbar/2$. The orbiting direction is determined by the spin-orbit interaction.

(7) Phase coherence is a consequence of the fact that the $2\lambda_L$ orbits are strongly overlapping.

(8) In the presence of a magnetic field $H$, the velocity of one of the members of the Cooper pair increases by
$e\lambda_L H/(m_e c)$ , the other decreases by the same amount,
and superconductivity breaks down when the slower velocity goes to zero, which determines the critical field as
\beq
H_{crit}=\frac{\hbar c}{4 e \lambda_L^2}
\eeq
This is essentially the same as the lower critical field value $H_{c1}$ in BCS theory \cite{tinkham}.

(9) The charge distribution in the ground state of superconductors  is macroscopically inhomogeneous, with more negative charge
near the surface and  more positive charge in the interior \cite{chargeexp,electrodyn}. Even though this costs potential energy, it minimizes the total energy, i.e.
potential plus kinetic. The excess charge density near the surface is \cite{electrospin}
\beq
\rho_-= en_s\frac{v_{\sigma}^0}{c}
\eeq
and associated with it there is an electric field in the interior of superconductors pointing radially outward, that reaches maximum value
\beq
E_m=\frac{\hbar c}{4 e \lambda_L^2}=H_{crit}
\eeq
within a London penetration depth of the surface of the superconductor.

(10) Electrodynamics of superconductors is somewhat different from what is described by London theory  \cite{electrodyn}.
While London's equation in the form Eq. (14) still holds, the gauge obeyed by the vector potential $\vec{A}$ in that
equation is the Lorenz gauge rather than the Coulomb gauge. Electrodynamic equations are relativistically covariant,
given in terms of 4-vectors by
\beq
J-J_0=-\frac{c}{4\pi \lambda_L^2}(A-A_0)
\eeq
with
$J=(\vec{J}(\vec{r},t),ic\rho (\vec{r},t))$, $A=(\vec{A}(\vec{r},t),i\phi (\vec{r},t))$, $J_0=(0,ic\rho_0)$, $A_0=(\vec{A}(0,i\phi_0 (\vec{r},t))$, 
with $\phi$ the electric potential and
\beq
\phi_0(\vec{r})=\int_V d^3 r'         \frac{\rho_0}{|\vec{r}-\vec{r}'|} .
\eeq
In particular, this implies that applied electrostatic fields are screened over a distance given by the London penetration
depth rather than the much shorter Thomas Fermi length.
The value of $\rho_0$ is determined by the
  value of the charge density near the surface of the sample Eq. (96) and charge neutrality.

(11) There is a macroscopic spin current in the ground state of superconductors flowing within a London penetration 
depth of the surface \cite{sm}, which is a  macroscopic zero
point motion of the superfluid. It results from the expansion of orbits to radius $2\lambda_L$, or equivalently the 
outward motion of negative charge during the transition,  and
the spin-orbit interaction with the background positive charge distribution. 
We have called this the `Spin Meissner effect' \cite{sm}.
The electrodynamic equations for the charge and spin sectors are given in Ref. \cite{electrospin}.
%

(12) The pairing interaction and resulting critical temperature $T_c$ is largest for systems with holes conducting through
closely spaced negatively charged anions \cite{matmech}. The mass of the ions is not important to determine the value of $T_c$, even though a small isotope
effect can arise from small variations in the value of the ionic mass.

These and many more details of the theory are given in the references in \cite{holesc}.

\section{ Precedents and other  views of the Meissner effect}
Various elements of the `reductionist' view of the Meissner effect discussed in this paper, that are not part of the `emergent' BCS view, were 
in fact proposed  in early work before the 
theory of hole superconductivity was developed \cite{transitionprocess}. We list them here.

(1) H. G. Smith \cite{smith1,smith2}. proposed in 1935 that the perfect diamagnetism of superconductors results from electronic orbits that are 
very large compared with atomic dimensions. More specifically, Slater proposed in 1937 \cite{slater} that  {\it ``to produce superconductivity the orbits must be of the order of magnitude of 137 atomic diameters''}. 
This is closely related to the $2\lambda_L$ orbits discussed earlier. With $137=\hbar c/e^2$, atomic diameter
$2a_0$ with $a_0$ the Bohr radius, and atomic density $1/(2a_0)^3$, the radius of  Slater's  orbits is $2\lambda_L\times 
\sqrt{\pi/2}$, with $\lambda_L$ given by Eq. (11).

(2) K. M. Koch proposed in 1940-45 \cite{koch1,koch2,kochjusti} as an explanation of the Meissner effect  that electrons flowing  outward
in the transition to superconductivity, due to a thermoelectric effect or other reasons, would be deflected by the Lorentz force
in the azimuthal direction so as  to suppress the initial field,   just as discussed here.

(3) W. H. Cherry proposed in 1960 \cite{cg}  that in the transition to superconductivity there would be
 nucleation and growth processes where outward diffusion of electrons
would give rise under the action of the Lorentz force to currents  that would weaken the magnetic field in the
interior, and  electrons backflowing to compensate for the charge imbalance would transfer
the thrusts they receive from the magnetic field to the lattice. This is closely related to 
the processes  of radial outflow and backflow discussed here within the `reductionist view'.

(4) Several workers before BCS proposed models with `spontaneous currents', or `circulating currents'
in the ground state of superconductors. Among them  Bloch \cite{blochstrom}, Frenkel \cite{frenkel}, Landau \cite{landau}, Smith \cite{smith1,smith2}, Born, Cheng \cite{borncheng} and  Heisenberg \cite{heisstrom}.  The currents were assumed to be charge currents, with different orientations in different
domains so that no net magnetic field would be generated.
This is related to the spin currents in the ground state of superconductors resulting from $2\lambda_L$ orbits  
mentioned earlier \cite{sm}.

(5) F. London remarked in the Introduction to his 1950 book \cite{londonbook}     {\it ``According to quantum
theory the most stable state of any system is $not$ a   state of {\it static equilibrium}
in the configuration of lowest potential energy. It is rather a
kind of {\it kinetic equilibrium} for the so-called zero point motion, which
may roughly be characterized as defined by the minimum average total (potential + kinetic) energy''}. And he added 
  {\it ``It is not necessarily a configuration
close to the minimum of the potential energy (lattice order) which is the
most advantageous one for the energy balance, since by virtue of the
uncertainty relation the kinetic energy also comes into play. If the
resultant forces are {\it sufficiently weak} and act between {\it sufficiently light}
particles, then the structure possessing the smallest total energy would
be characterized by a good economy of the kinetic energy ... It would be the
outcome of a quantum mechanism of macroscopic scale.''}.
These concepts are not part of BCS theory, and foreshadowed the predictions of the theory of
hole superconductivity that  the ground state charge distribution in superconductors is macroscopically inhomogeneous, costing potential energy, and the 
energy lowering is due to quantum kinetic energy.

None of the physics discussed in (1)-(5) above is part of the conventional BCS `emergent viewpoint' of the
Meissner effect, and all of it is related to the `reductionist viewpoint' discussed here.

Finally, we mention here  some other viewpoints on the Meissner effect that have been proposed.
We don't  believe they are right.

(1) Essen and Fiolhais \cite{essen} proposed in 2012 that magnetic flux expulsion from a sample cooled to superconductivity can be understood as an approach to the magnetostatic energy minimum, with 
quantum mechanics playing no role, and without the necessity of a condensation energy for
the superconducting state.

(2) Kr\"uger \cite{kruger} in 2017 proposed that the Meissner effect results from `quantum
mechanical constraining forces' that in the presence of a magnetic field produce Cooper pairs
of non-vanishing total momentum in nearly half-filled narrow energy
bands.

(3) Nikulov \cite{nikulov1,nikulov2} argued in 2022 that in the superconductor to normal transition the supercurrent
stops through normal scattering processes generating Joule heat,
and that this implies that the Meissner effect is a pheonomenon where the second law
of thermodynamics is violated.

(4) Koizumi \cite{koizumi} proposed in 2024 that there are several 
additional important forces besides the Lorentz force  in superconductors, in particular there is a 
`force' for producing topologically protected loop currents that explains the Meissner effect.
 
(5) Kozhevnikov \cite{koz} proposed in (2021) that 
{\it ``a novel semi-classical micro-whirls model''} explains the Meissner effect.
Whether or not this has any relation with the $2\lambda_L$ orbits discussed here was not discussed in  \cite{koz}.

 \section{A test case}
 Here we discuss a test that may help decide which  of  the two different points of view on the Meissner effect
 discussed here is correct.
 
 Consider a simply connected superconductor with a small cavity totally enclosed in its interior, as shown in Figs. 22 and 23. We show a spherical body with a spherical cavity, but the discussion here
 should apply to any simply connected body with a small cavity of any shape that is completely
 enclosed withing the body. We assume the dimensions of the cavity are much
 smaller than the dimensions of the body.

 The system is cooled in the presence of a small magnetic field. Will the field be completely expelled?
 
 The state shown on the right panel of Fig. 22 has currents circulating only near the outer surface
 of the body and the magnetic field is completely expelled. 
 This is the lowest energy state of the system. 
 
 Instead, the state shown on the right panel of Fig. 23 has the same currents circulating near
 the outer surface of the body as in Fig. 22, and additionally has currents circulating around the cavity
 and around a cylindrical region that extends from the bottom to the top of the body.
 In this cylindrical region the system is in the normal state. This state is $not$ the lowest energy state of the system.
 
 In the state shown on the right panel of Fig. 23 the magnetic field is trapped in the
 cavity and the cylindrical regions above and  below it. The system cannot evolve to the lowest energy state
 shown in Fig. 22 right panel because this would require magnetic field lines
 to move across superconducting regions, which is not allowed. That is clear.
 
 It is also clear that the state shown on the right panel of Fig. 22 can be reached
 by first cooling the system into the superconducting state in the absence of
 magnetic field and then applying the magnetic field. That is the same situation discussed in
 Sect. VIII, Fig. 3 panels {\bf a} and {\bf b}.

               \begin{figure} []
 \resizebox{6.5cm}{!}{\includegraphics[width=6cm]{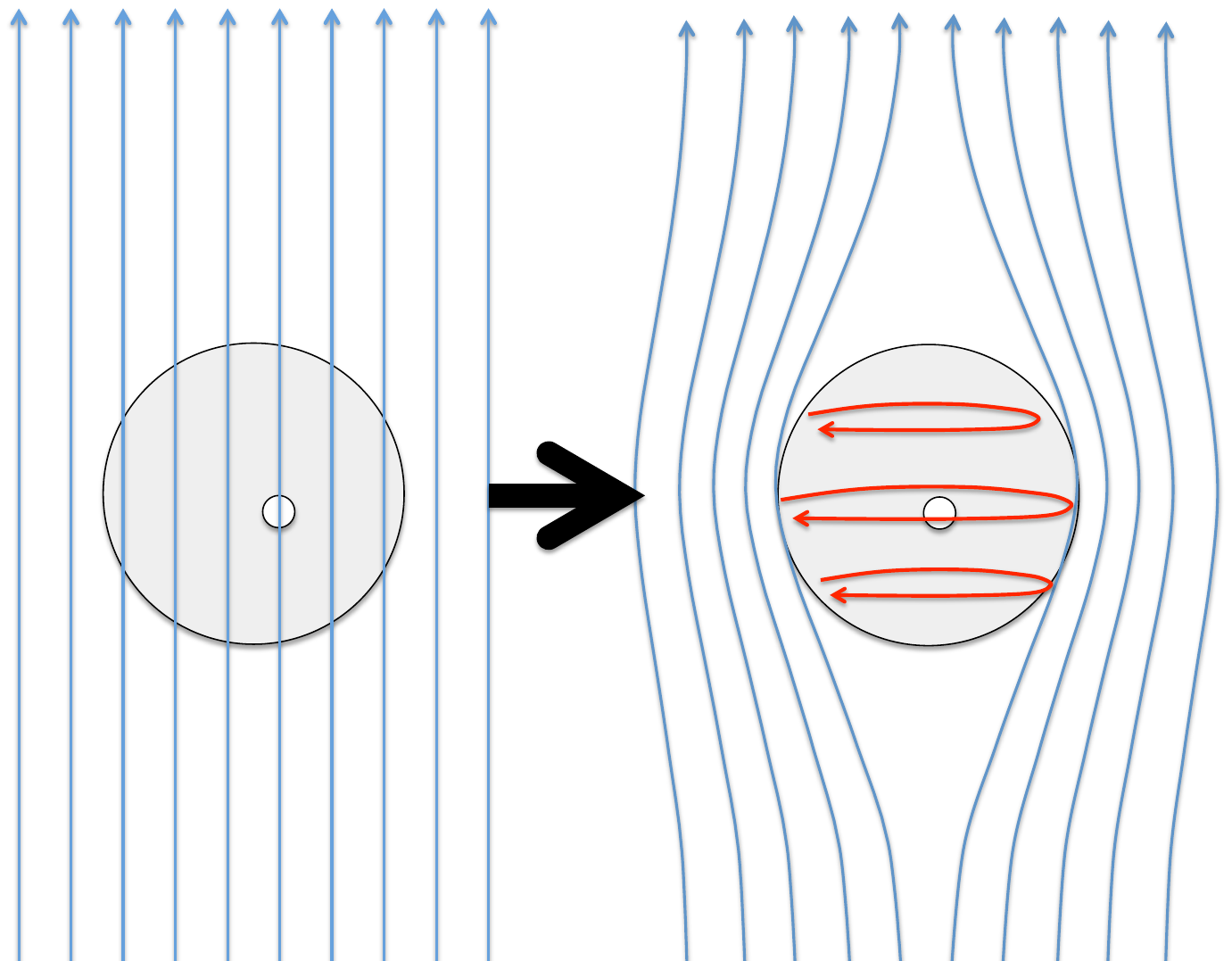}} 
 \caption {Cooling of a simply connected superconductor with a small cavity in its interior in a small uniform magnetic
 field (blue lines).
 The state of lowest energy has currents (red curves) circulating near the surface only and the magnetic
 field is totally expelled from the interior including the cavity.   }
  \label{figure1}
 \end{figure}

 What is not clear \cite{poole}  is whether the  state shown on the right panel of Fig. 22 can ever be
 achieved under any cooling protocol, starting in the normal state in the presence of a 
 magnetic field. One could try different cooling protocols \cite{trappingrate}, for example lowering
 the temperature first near the center of the body, or in the region near the cavity,
 or near the surface of the body; cooling at different rates, for example lowering the
 temperature very quickly so the system would supercool to a temperature much
 lower than $T_1$ with the applied $H=H_c(T_1)$. Or cooling very slowly
hoping to  avoid that the system gets stuck in metastable states. 
 Or combining inhomogeneous temperature distribution patterns and rates of cooling in
 a variety of ways.
 
 We argue that within the `emergent point of view' discussed in this paper, there is no
 reason why there could not be a cooling protocol that would lead the system to evolve
 from the initial state on the left panel of Fig. 22 to the final state on the right panel of
 Fig. 22, which is the lowest energy state of the system. It is generally believed within
 that point of view that systems will ``find a way'' to reach their lowest energy state,
 as discussed earlier. The system shown in Fig. 22 right panel is
 described by a macroscopic wavefunction $\psi(\vec{r})=n_s^{1/2}(\vec{r})   e^{i\theta(\vec{r})}$ that is macroscopically
 phase coherent, the phase of $\psi(\vec{r})$ is the same at every point in the material. Of course
 inside the cavity the phase is undefined because there is no material.
 There is nothing in BCS theory, Ginzburg-Landau theory,
 time-dependent Ginzburg-Landau theory, nor the Higgs mechanism, that says that
 such a macroscopic quantum state cannot emerge in such a many-body system with
 quintillions of interacting particles, given that  `more is different' \cite{emergent,anderson,coleman}.
 Minimization of the Ginzburg-Landau free energy of the system shown in Fig. 22 certainly leads to 
 the state shown on the right panel of Fig. 22 and not to the one on the right panel of Fig. 23.

   \begin{figure} [t]
 \resizebox{6.5cm}{!}{\includegraphics[width=6cm]{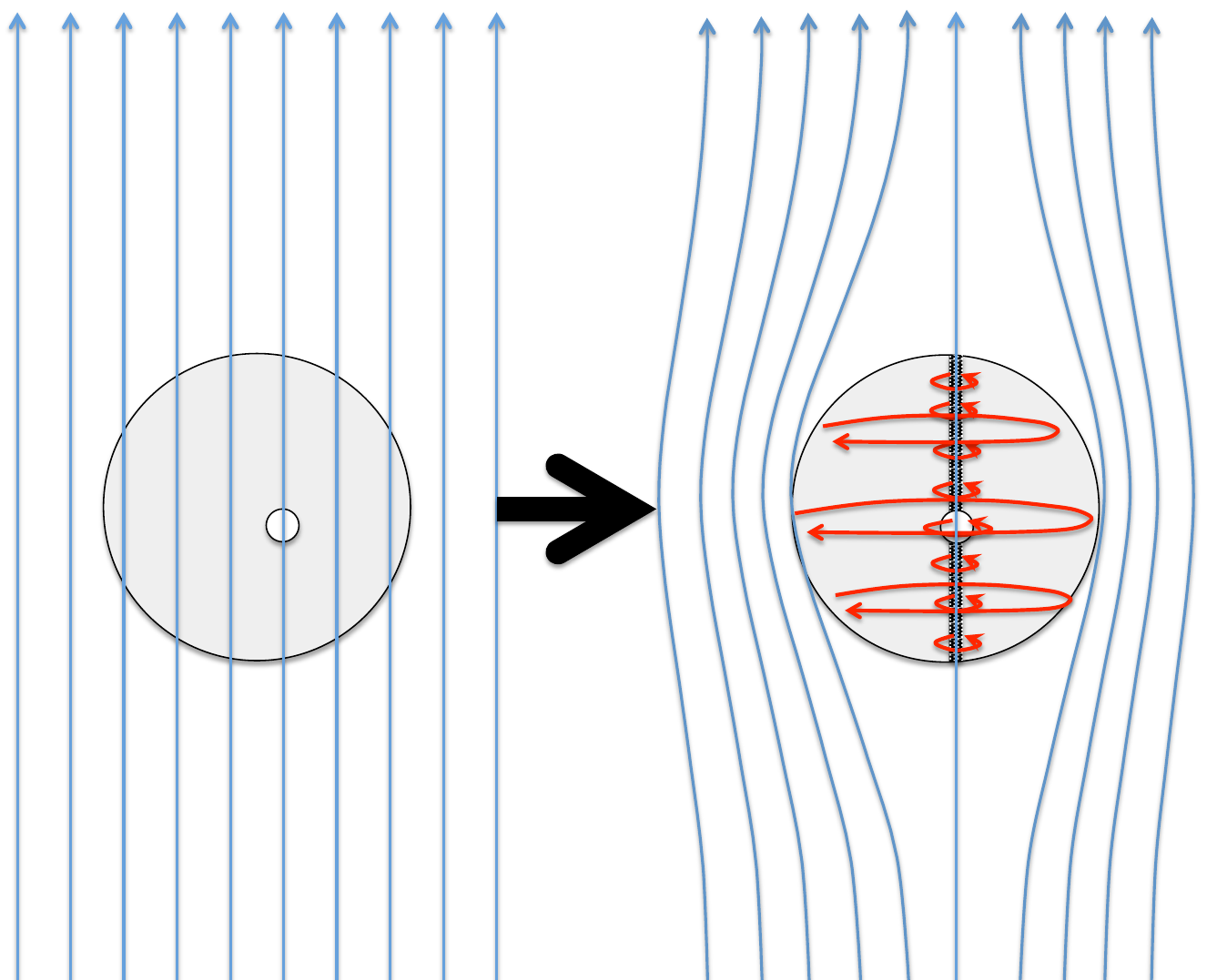}} 
 \caption {A possible outcome of cooling a simply connected superconductor with a small cavity in its interior in a magnetic
 field.
The state shown on the
 right panel has magnetic flux trapped and higher energy than the state shown in Fig. 22
 right panel. }
  \label{figure1}
 \end{figure}

 Instead, within the `reductionist point of view' discussed in this paper, the final
 state will necessarily be the `metastable state'  shown on Fig. 23 right panel, 
 the lowest energy state of Fig. 22 right panel cannot  be reached starting from the system in the normal state with
 magnetic field in the interior \cite{meissnerlocal}.
 In essence, this is simply because within that point of view expulsion of magnetic field requires
 radial charge flow, and there cannot be radial charge flow
 inside the cavity because there is no material there. If the magnetic field
 cannot be expelled from the cavity those magnetic field lines have to 
 carve a path through the material to reach the outside of the body, paying the energy cost of destroying the
 superconductivity in their path. 
 
 An interesting consequence of this physics predicted by the reductionist
 viewpoint \cite{scstops} is  that the reverse process to that shown in Fig. 22,  i.e. from the right panel to the left panel, 
will necessarily involve dissipation of some 
 Joule heat, no matter how slowly it takes place, simply because
 the process is $irreversible$. Instead, within the `emergent' viewpoint the process is reversible, so no
 Joule heat would be dissipated in either direction if the transitions occur infinitely slowly.
   
 We note  that these tests could also be performed with a rapidly rotating body that is cooled into the 
 superconducting state in the absence of applied fields. If the `emergent' point of view is correct,
 the resulting magnetic field should be perfectly uniform in the interior of the body including the cavity
 and given by the value Eq. (65). For a spinning sphere, the resulting magnetic moment (`London moment' \cite{londonbook})
 would be precisely 
 \beq
 m=\frac{m_e c}{e}R^3 \omega 
 \eeq
for a sphere of radius $R$ spinning with angular velocity $\omega$, irrespective of whether the sphere has a cavity in its interior or not. 
 Instead, if the `reductionist' point of view is correct,
 the magnetic field should be suppressed in a cylindrical region that goes through the cavity, and the resulting
 London moment whould be smaller than Eq. (100), and smaller than what would result if 
 the sphere is first cooled and then put into spinning motion, which would be Eq. (100).
 
 In summary, experimental tests such as discussed in this section
 and similar ones \cite{radialflow}  could rule out the
 `reductionist viewpoint' if it is found that processes such as the one  shown in Fig. 22 can take place in nature.
 Or, they could lend support to it if it is found that they cannot.

\section{ Summary,   conclusions and implications}

In this paper we have reviewed two different ways to understand the Meissner effect, the most fundamental property of superconductors.

In one way, which we called the `emergent  point of view', which is the overwhelmingly generally accepted point of view, there is no need to understand in detail the processes involved in the Meissner effect.
Given a theory that describes the superconducting state with magnetic field excluded as the lowest energy state of the system, it is postulated that
the system will find a way to reach this lowest energy state. 
The Meissner effect is a $consequence$ of this fact. 
Within this point of view, there is no point in asking further questions \cite{bcs50,needed}. The conventional theory of superconductivity is such a theory.

However, we argued that there is experimental evidence that in our view casts strong doubt on the validity of this point of view.
Namely, the fact that it is {\it extremely easy} for systems to $not$ completely expel magnetic fields . 
In the real world the degree of magnetic field expulsion is extremely dependent on the purity of  samples and is never complete, even 
for extremely pure samples  \cite{mend,shoen,cavities}. This flies in the face of the article of faith that systems know how to reach their lowest energy state: in the presence
of impurities or defects, magnetic field remains trapped, turning parts of the sample that would prefer to be superconducting
normal, which costs energy compared to the state where the entire field is excluded \cite{meissnerlocal}. Why doesn't the system find its way to 
its lowest energy state?

In an alternative way to understand the Meissner effect, which we called the `reductionist point of view', exposed within the theory of hole superconductivity \cite{holesc},
it is thought that it is necessary to understand the dynamics of the
processes involved, which forces cause which processes, how momentum transfers between different parts of the system occur, what constraints are imposed by the fact that systems need to obey
conservation laws and the laws of thermodynamics, what constraints are imposed by Bohr's correspondence 
principle and the principles of magnetohydrodynamics, and what physical elements should the system possess to allow it to 
undergo  these processes.  This understanding poses constraints on 
which systems can be superconducting and which cannot. In particular, it says that if the system does not have
normal carriers with negative effective mass at the Fermi energy, it cannot reversibly interchange momentum between 
electrons and the body as a whole, which is necessary for the Meissner effect and its reverse to occur,
hence such a system would not be a superconductor.

The crucial difference between both points of view, which should be amenable to experimental verification, is whether or not 
there is radial charge flow associated with the motion of magnetic field lines in the Meissner effect.
Within the emergent point of view there isn't, within the reductionist point of view there is. 
Within the emergent point of view, charge carriers acquire azimuthal momentum spontaneously in the process of
nucleation and growth of the superconducting phase in the presence of a magnetic field. 
Within the reductionist point of view, azimuthal momentum is a result of radial momentum plus the Lorentz force.
As discussed in this paper, the mechanical momentum that needs to be transferred between electrons and ions
during the Meissner effect is many orders of magnitude larger than the physical momentum of the supercurrent in the
final state, so to understand how it happens is essential.

For  classical conducting fluids, magnetohydrodynamics dictates that motion of magnetic field lines is closely associated with motion 
of the fluid. Within the emergent point of view, that physics is irrelevant to the Meissner effect. Within the reductionist point of view,
it plays a key role.

We have pointed out that the Meissner effect and its reverse are understood to be {\it reversible processes} in an ideal situation,
which implies that the entropy of the universe should not change in the processes \cite{reversible}. This  requires an explanation of how 
mechanical momentum
is transferred between electrons and ions in a reversible way, since the Meissner current carries mechanical momentum.
Within the reductionist approach discussed here, transfer of momentum between electrons and ions is mediated by the electromagnetic
field, and no scattering processes are involved. For this to be possible requires radial motion of charge, to give rise to
radial electric fields and associated electromagnetic azimuthal momentum. Within the emergent approach there is no radial motion
of charge, which implies that there is no azimuthal electromagnetic field momentum, which implies  that the electromagnetic field does not mediate the momentum transfer between electrons and ions.
How such momentum transfer occurs without generation of entropy within the emergent point of view requires an explanation.

Within the emergent point of view, the physical process by which a quantum particle  spontaneously acquires azimuthal momentum
in opposition to the Faraday electric force acting on it 
is not specified, it 
would be something unique to the superconducting transition. A. Nikulov has attempted to provide
a physical justification for it by postulating that there is an azimuthal ``quantum force'' that acts
in the process \cite{nikulovqf,nikulovqf2}. This quantum force would be a new force of nature that acts in no
other physical situation and has no classical   counterpart. In addition, it requires that there is another quantum force that is equal and opposite
and acts azimuthally on the ions, so that momentum is conserved.  We argue 
that these requirements make this  point of view highly  implausible.

   \begin{figure} [t]
 \resizebox{6.5cm}{!}{\includegraphics[width=6cm]{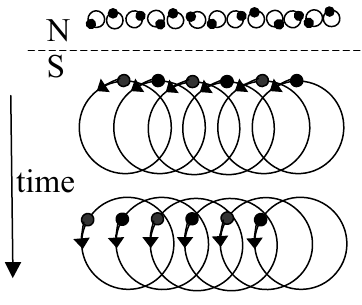}} 
 \caption { Schematic explanation for why phase coherence is required in the superconducting phase (S)  due to expansion of the orbits discussed in the text and not in the normal (N) phase. The black dots represent the position of the electron in its orbit, and its angular
 position is
 the `phase'. }
  \label{figure1}
 \end{figure}

 Within the reductionist  point of view, a physical process by which a quantum particle spontaneously acquires radial 
momentum is not difficult to conceive. Quantum particles lower their quantum kinetic energy by
expanding their wavefunction \cite{emf}, so they will do so if a constraint preventing it from happening is 
removed. The constraint that does not allow electrons in the normal phase to expand their wave function
is that the superconducting phase  requires {\it phase coherence}, because the resulting wavefunctions / orbits in the
many-electron system will overlap, as shown in Fig. 24. Above the transition temperature
the higher entropy of the normal phase, where each electron has its own individual random  phase, dominates,
and the free energy $F=E-TS$ is lowest. Below the transition temperature the superfluid electrons are willing
to pay the entropy cost entailed in all having the same quantum phase  required by the overlapping orbits (Fig. 24) because $TS$ is smaller when $T$ is lower, and the lowering
of quantum kinetic energy achieved by wavefunction/orbit expansion dominates. Therefore, orbits will expand and overlap, driven by a real force, which originates in quantum pressure. Quantum pressure yields a  force pointing radially out,
$-\partial / \partial r (\hbar^2/(2m_e r^2)$,  that overcomes the 
radially inward force of the Maxwell pressure,  it is the `Meissner pressure' hypothesized by London \cite{londonbook}.
Instead, it is difficult to conceive how an attractive interaction that leads to the lowering of $potential$ energy
upon pairing, as assumed within the conventional theory of superconductivity, would give rise to an outward
pressure that overcomes the Maxwell pressure.

%
%
 

  If the first point of view discussed here (emergent)  is right, the second point of view is wrong. If the second point of view discussed here (reductionist)
  is right, the first point of view is wrong.
 There is no room for compromise.  
 
What are the larger implications of this?  If the emergent point of view is correct, it poses no constraints on mechanisms giving rise to superconductivity.
 The electron-phonon interaction for conventional superconductors, and any of the myriad of pairing mechanisms that have been proposed for 
 unconventional superconductors arising from correlated electron physics, remain plausible candidates to 
 explain the superconductivity of known materials \cite{specialissue} and materials to be discovered.
 Within this prevailing point of view, the most promising way to achieve high temperature superconductivity is with materials with light atoms,
 and much effort is currently  focused in that direction \cite{roadmap}.
 
 If the reductionist point of view discussed here  is correct, it poses strict constraints on mechanisms of superconductivity. It implies that the electron-phonon interaction
 is not the mechanism giving rise to superconductivity in any material \cite{xor,apl}, instead superconductivity originates in electron-hole
 asymmetric electron-electron interactions that lower the kinetic energy of carriers upon pairing and requires that the normal
 state carriers have hole-like character. Within that point of view,  the most promising way to achieve high temperature superconductivity is with materials containing negative anions in close proximity with hole conduction through them, like the high $T_c$ cuprates and $MgB_2$.

Therefore, for the physics community to determine what is the correct way  to understand the Meissner effect in superconductors is urgent and important.
  

\end{document}